\pdfoutput=1

\documentclass[a4paper,11pt]{article}
\usepackage{jheppub}
\usepackage{aas_macros}

\usepackage{graphicx}
\usepackage{placeins}

\usepackage{xcolor}
\usepackage{amsmath}
\usepackage{amssymb}
\usepackage{bm}
\usepackage{slashed}
\usepackage{xspace}
\usepackage{makecell}  
\usepackage{adjustbox} 
\usepackage{enumitem}
\usepackage[normalem]{ulem}

\usepackage{fontawesome5}
\usepackage{hyperref}
\usepackage{orcidlink}

\usepackage{multirow}
\usepackage{array} 
\newcolumntype{L}[1]{>{\raggedright\let\newline\\\arraybackslash\hspace{0pt}}m{#1}}
\newcolumntype{C}[1]{>{\centering\let\newline\\\arraybackslash\hspace{0pt}}m{#1}}
\newcolumntype{R}[1]{>{\raggedleft\let\newline\\\arraybackslash\hspace{0pt}}m{#1}}

\title{Detecting White Dwarf Binary Mergers with Gravitational Waves}

\author[a]{Giona~Sala\orcidlink{0009-0001-3716-862X},}
\affiliation[a]{Institute for Theoretical Particle Physics and Cosmology, RWTH Aachen University,\\D-52056 Aachen, Germany}
\emailAdd{gsala@physik.rwth-aachen.de}

\author[b]{Chiara~Brandenstein\orcidlink{0009-0004-9538-4795},}
\affiliation[b]{Leinweber Institute for Theoretical Physics at Stanford, Stanford University, Stanford, CA 94305, USA}
 \emailAdd{chiaramb@stanford.edu}

\author[a, c]{Sebastian~Baum,}
\affiliation[c]{Mid Sweden University, 831 25 \"Ostersund, Sweden}

\author[b,d]{and Peter~W.~Graham}
\affiliation[d]{Kavli Institute for Particle Astrophysics and Cosmology, Department of Physics,\\
Stanford University, Stanford, CA 94305, USA}

\abstract{Mergers of white dwarf binaries are a possible progenitor channel for Type Ia supernovae. While white dwarfs are abundant in the universe and relatively well understood, their gravitational wave signals have not yet been directly observed.  In order to detect gravitational waves from merging white dwarf binaries, a detector in the mid-band between LVK and LISA appears necessary. In this paper, we compute and discuss the gravitational waves emitted by inspiraling and merging white dwarf binaries, and assess their detectability with proposed space-based atom-interferometer detectors such as MAGIS Space and AEDGE. 
Gravitational waves from massive white dwarf binaries can be observed for many years before merger, offering a unique early warning of their final explosion.
Our projections suggest that MAGIS Space could detect signals from Type Ia supernova progenitors at least once every four years, while AEDGE could observe at least a few hundred such events annually.
The prolonged gravitational wave emission captured by atom-interferometers provides precise sky localisation and can allow observation of the final explosion with electromagnetic telescopes.
The combined observation with electromagnetic radiation from the white dwarf binary coalescence could open a new pathway for multi-messenger astronomy involving some of the brightest transient events in the universe.}

\begin{document} 
\maketitle
\flushbottom

\section{Introduction} \label{sec:Intro}

White dwarfs (WDs) are compact remnants of stars that maintain structural stability through a balance between gravitational forces and electron degeneracy pressure.
They are among the most common astrophysical objects in our galaxy, and have been extensively studied during the last century~\cite{Chandrasekhar_1931,Shapiro_1983,Holberg_2009}. Many of these compact objects are expected to be found in pairs~\cite{Marsh_1995,Nelemans_2001,Toonen_2012, Lamberts_2019}; however, white dwarf binaries (WDBs) remain difficult to detect because of their faint electromagnetic (EM) emission and challenges in differentiating WDBs from single WDs~\cite{Ira_2017,Toonen_2017, Burdge_2020, Burdge_2025}. 
Gravitational wave (GW) signals from WDBs have never been directly measured, but they are generating considerable interest in the scientific community~\cite{Ruiter_2010,Maselli_2020}. In fact, they could constitute one of the primary sources for future GW detectors operating within the micro- and milli-Hertz frequency bands~\cite{Seppe24,Hofman24}.

\begin{figure}[t]
    \centering
    \includegraphics[width=\textwidth]{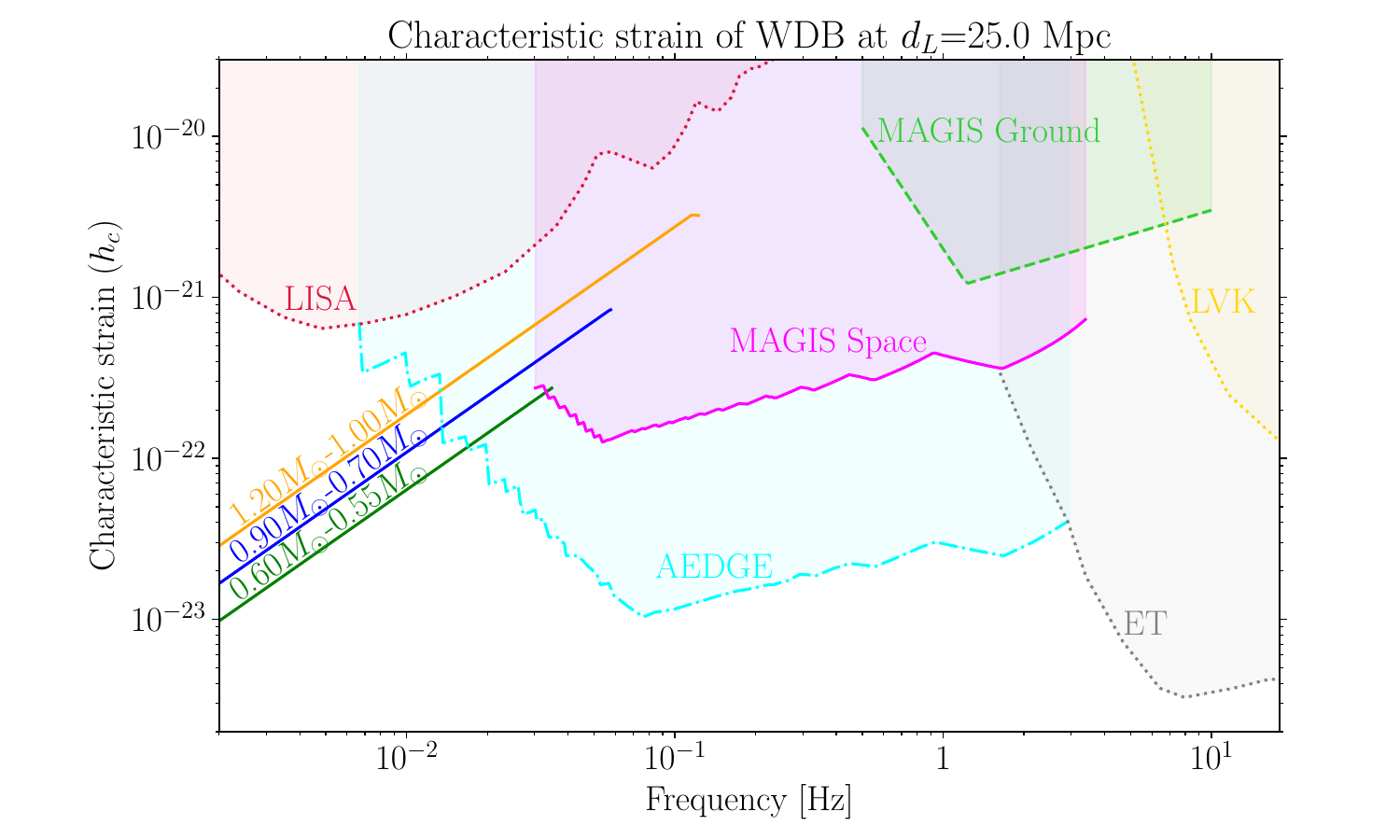}
    \caption{Characteristic strain sensitivity of the main GW detectors, currently operating (LVK), in development (LISA and MAGIS Ground) and proposed (MAGIS Space and AEDGE), in the frequency range of interest~\cite{MAGIS-100:2021etm, AGIS, AEDGE, Graham:2016plp}. The signal from three representative WDBs, labelled directly on the plot, for $1$\,yr of observation, or until they start to mass transfer at Roche Lobe Overflow (see \autoref{sec:RLOF}),  is shown in the plot. A distance of $25.0$\,Mpc is chosen, as it represents the highest contribution region to the detectable GW signal that might go SN Ia (see \autoref{sec:GW}). }
    \label{fig:CharacteristicStrain}
\end{figure}

Gravitational wave detectors in the mid-band, ranging from roughly $10$\,mHz to $10$\,Hz, could observe signals of merging WDBs. The observability of the binary depends on the component WD masses and sizes, which determine the frequency at coalescence~\cite{Kinugawa_2022,Mandel:2017pzd, Sedda:2019uro, Maselli_2020}. 
The inspiral of these compact objects occurs over very long time scales, making it possible to observe them over extended periods and gain deeper insight into the underlying astrophysical processes.
The most violent and interesting mergers involve massive WDBs, which also means that the component WDs are more compact, thus the binaries reach higher frequencies.
More specifically, WDBs with masses $\gtrsim 0.6\,M_\odot$ merge within a few years after the start of  mass transfer, and happen at frequencies between $50$\,mHz and $1$\,Hz, below the sensitivity range of terrestrial laser interferometers such as the LIGO–Virgo–KAGRA (LVK) collaboration~\cite{Advanced_LVK}. Even next-generation detectors like Einstein Telescope (ET)~\cite{ET} and Cosmic Explorer (CE)~\cite{CE} are only sensitive down to a few hertz. 
On the other end, pulsar timing arrays~\cite{NANOGrav:2020bcs,NANOGrav:2023gor} probe the nano-hertz regime, whose frequencies are much lower than the WDB merger signal. 
LISA will see many WDBs before their coalescence, but its sensitivity drops right before the frequencies of interest, preventing  observation of  merging WDs~\cite{LISA,LISA:2017pwj,LISA:2022yao,toubiana2024interacting}. 
On the other hand, new mid-band GW detectors will be able to tackle this challenge. 
Some of the proposed instruments that are expected to measure WDB mergers are MAGIS Space~\cite{AGIS}, AEDGE~\cite{AEDGE}, and the Lunar Gravitational Wave Antenna (LGWA)~\cite{lunar_2024}. In the article, we focus on MAGIS Space, and secondarily, also on AEDGE.
The various sensitivities are graphically shown in \autoref{fig:CharacteristicStrain}. As can be seen in this figure, ground based atom interferometers will not be sensitive to WDB mergers, because they are not able to measure at low enough frequencies due primarily to gravity gradient noise.

Detecting the GW signal from WDB coalescences could open a new window for multi-messenger astronomy by enabling the prediction and subsequent observation of the EM counterpart, potentially a  Type Ia supernova (SN Ia)~\cite{Liu:2023qmw}. 
The process of two WDs detonating into a SN Ia is called the double degenerate (DD) scenario, as opposed to the single degenerate (SD) channel, for which a WD accretes mass from a different companion.
The DD mechanism describes the process of two rather heavy WDs merging and exploding as a SN Ia~\cite{Hillebrandt:2013gna, Maoz_2014}, and represents a promising progenitor of certain SN Ia~\cite{Livio:2018rue}.
The measurement of both GW and EM radiation from a single event can lead to a deeper understanding of SN Ia, which is of great interest for advancing our knowledge of astrophysics and cosmology.
In particular, SN Ia are reliable standard candles, playing a crucial role in measuring distances in the late universe. Their observation can be used to accurately constrain key cosmological parameters.
The absence of a SN Ia signal after the merger of two massive WDs, or the observation of alternative signatures, would also be of significant interest, as it could reveal unknown late-stage dynamics and point to alternative astrophysical phenomena.

In this project, we estimate the rates and precision of detecting WDBs with atom-interferometers. 
We find that we can expect to measure roughly one merger on a biannual basis with MAGIS Space, and more than a hundred yearly with AEDGE. Around half of these events are expected to produce visible SN Ia. 
We can expect that all of these detonations can be observed electromagnetically. On the one hand, we find the moment of merger can be recognised promptly: for low mass WDB, the final burst can be forecasted well in advance because of the slow evolution; for more massive objects, predicting the quick disruption of the binary is hard, but the disappearance of the GW signal can be identified very swiftly. It is possible to notice the absence of WDB in time to study the SN Ia afterglow, which lasts for months.
On the other hand, sky localization in this band is excellent~\cite{Graham:2017lmg, Baum:2023rwc}.  We find that for WDBs with appropriate masses to detonate as SN Ia at the end of their life time, MAGIS Space could reconstruct the sky position of a binary at a distance of 25\,Mpc with uncertainty as small as $10^{-2}$\,${\rm deg}^2$.  
Consequently, one could expect the observation of a multi-messenger event at least every $4$\,years with MAGIS Space. AEDGE would lead to hundreds of multi-messenger measurements per annum.
The result of a similar analysis for the LGWA concept~\cite{marcano2025massivedoublewhitedwarf, lunar_2025} falls quite nicely within these ranges. 

In \autoref{sec:Detector} we introduce our probe of interest, MAGIS Space, illustrate its experimental setup and discuss its detection capabilities. In the following \autoref{sec:WDBevo}, we describe WDBs and characterise their main properties, which are relevant for our analysis. In \autoref{sec:GW} we explain our methods and computation. More specifically, we introduce and infer the precision on different GW parameters. Following up, in \autoref{sec:Mergersignal} we discuss how well a WDB merger can be identified and SN Ia can be predicted. The results on the expected merger rates and consequent SN Ia rates are shown in \autoref{sec:mergerrate}.
We conclude the article with a summary of the main findings in \autoref{sec:Discussion}. Furthermore, 
\autoref{app:corrections} provides a summary of the main assumptions and limitations made in this work.

\section{Atom-Interferometers as GW Detectors in the Mid-Band} \label{sec:Detector} 

Gravitational waves are challenging to detect due to their extremely weak signals. Currently, laser interferometers like those in the LIGO–Virgo–KAGRA (LVK) collaboration enable the observation of gravitational waves from neutron star and black hole mergers with frequencies ranging from approximately $10$\,Hz to $10$\,kHz~\cite{PhysRevLett.116.061102, PhysRevLett.119.161101,LIGOScientific:2016vlm,LIGOScientific:2018mvr,LIGOScientific:2020ibl,LIGOScientific:2021usb,KAGRA:2021vkt}. 
Also, pulsar timing arrays (PTA) have detected gravitational waves~\cite{nanograv} in the nHz band, whose source is still debated. 
For GW searches, the different frequency bands are complementary and should all be explored. 
Different GW frequency bands allow observation of sources from a wide variety of astrophysical and cosmological origins, which could be unveiled and studied by a broad spectrum analysis of GWs, similar to observations of astrophysical phenomena in the EM spectrum from radio waves to gamma rays. A combined effort of multiple instruments to cover the whole spectrum is required.
Currently, there is a gap in the observational coverage in the so-called mid-band, between $10$\,mHz and $10$\,Hz. 
LISA will be a laser interferometer, with a configuration similar to LVK detectors, but operating in space, and will be most sensitive in the frequency range approximately between $1$\,mHz and $50$\,mHz~\cite{LISA,LISA:2017pwj,LISA:2022yao}.
In between LIGO and LISA, atom-interferometer experiments are one promising possibility to close the remaining gap, see \autoref{fig:CharacteristicStrain}.
The mid-band is particularly interesting because many astrophysical sources can be observed from inspiral to merger and ringdown.  Seismic, thermal and other displacement noise limits ground based interferometers like LVK. Atom interferometers can escape some of that noise by using free-falling atoms, but they are still subjected to gravity gradient noise. Space-based setups face less gravity gradient noise and thus can detect signals at much lower frequencies \cite{PhysRevD.111.082003}.
These atom-interferometers will then be sensitive in the mid-band, more specifically between roughly $10$\,mHz and $10$\,Hz, enabling the study of many compact binaries including WD binaries with component masses down to $\sim 0.1\,M_{\odot}$. 
At the same time, such mid-band detectors can pinpoint the location of GW sources in the sky with sub-degree localisation~\cite{Graham:2017lmg, Baum:2023rwc}, for observation durations between a few months and a year, if the compact binaries are long-lived and visible during this period. 

LVK and LISA detect gravitational waves using a Michelson interferometer, in which light serves as the source of interference. A laser beam is split by a partially reflecting mirror into two perpendicular paths, each directed toward a reflective “test mass”, the mirrors. After reflection, the beams are recombined, producing an interference pattern that encodes changes in the relative path lengths, such as those caused by passing GWs.
Atom interferometers follow the same basic principle, but the atoms' wavefunctions are split by ``mirrors" made by laser pulses.  However, even though the atoms are splitting, in the MAGIS concept~\cite{dimopoulos2008atomic, Dimopoulos_2009, Graham:2012sy, Graham:2016plp, Graham:2017pmn} the atoms are actually playing the role of the test masses, analogous to the mirrors in LIGO. Two clouds of ultra-cold atoms are placed in free fall on the ends of a long laser baseline, then subjected to laser pulses that place the atoms in a quantum superposition of two spatially separated paths. Subsequent pulses recombine these paths, and the difference in phase shifts measured in the two atom interferometers on the ends reveals tiny changes in the baseline distance induced by GWs. In both atom and optical interferometers, a differential phase measurement is made, so laser phase noise and rigid platform motion are suppressed. Gravity gradient noise does not cancel out since it is not common mode.

This concept is closely related to linking two atomic clocks with laser pulses, and measuring the light travel time between the clocks~\cite{Kolkowitz:2016wyg}. However, atomic clocks are highly sensitive to mechanical vibrations and thus an inertial reference system must be used. 
To overcome this, an atom interferometer uses free-falling neutral atoms as test masses, which largely decouples the measurement from environmental disturbances.  A common laser pulse is used to drive both atom interferometers, which suppresses laser phase noise.

Terrestrial versions of such atom interferometer GW detectors are now under construction, mostly 100-meter class demonstrators, including  MAGIS-100~\cite{MAGIS-100:2021etm}, AION~\cite{Badurina:2019hst}, MIGA~\cite{MIGA}, and ZAIGA~\cite{Zhan_2019}.
These will prove the technology as preparation for larger km-scale terrestrial versions.
Additionally, satellite atom interferometer detectors are planned, including MAGIS Space~\cite{AGIS,Graham:2012sy,Graham:2016plp,Graham:2017pmn} and  AEDGE~\cite{AEDGE}. These experiments will use two satellites, each with an atom-interferometer, and connect them via a long laser baseline. 
The two satellites of the MAGIS Space experiment are expected to have a geocentric medium Earth orbit (MEO) of radius around $2\times10^4$\,km, with a period of approximately $10$ hours, avoiding any interaction with Earth’s upper atmosphere~\cite{Graham:2017pmn}.
The planned sensitivities are shown in \autoref{fig:CharacteristicStrain}.
Note, for AEDGE the sensitivities are about an order of magnitude better. This is because the estimated phase sensitivity in the interferometer is one order of magnitude lower, which could be achieved with a higher atom flux (or a version of squeezing) \cite{El_Neaj_2020}. Also, for ground based interferometers, seismic, thermal and other noise sources dominate in the mid-band. Consequently,  these terrestrial detectors are not sensitive to the lower frequencies at which WDB's merge. These noise sources are eliminated once the experiment is performed in space.
The satellite experiments are most sensitive in the $0.1-1$\,Hz range, the band in which WDB mergers could be observed. By tuning the laser pulse sequence, both MAGIS Space and AEDGE can operate in a resonant mode. The interferometers can also perform a broadband search initially to locate the source and can be easily switched to resonant mode as soon as a source is identified.
This can allow WDBs to be observed at high signal-to-noise ratio (SNR) by tracking their frequency evolution.

By operating in this mid-frequency band around 0.01-1 Hz, these detectors will be able to measure the sky position of the source quite accurately~\cite{Graham:2017lmg, Baum:2023rwc}.  Roughly, this arises from looking at the phase lag of the signal (or change to arrival time) in the single detector, while the detector moves over a large distance (effectively a synthetic aperture). 
In our case, this is $\sim 1 \, \text{AU}$ as the satellite moves around the Sun (so long as the source lasts several months). This enables localising sources with sub-degree precision in the mid-band, where sources are inspiralling for long periods of time.

In this work, we investigate the capabilities of the MAGIS Space and AEDGE atom-interferometers.
We find that MAGIS Space is expected to detect numerous white dwarf binaries (WDBs) with component masses exceeding
$0.5\,M_{\odot}$, as illustrated in \autoref{fig:CharacteristicStrain}. A substantial fraction of these systems is anticipated to eventually merge and trigger SN Ia. This opens the prospect for multi-messenger observations, where gravitational wave detections from MAGIS Space can be correlated with electromagnetic signals from SN Ia, providing new insights into both binary evolution and supernova progenitors. In this paper, we explore the detector’s sensitivity to such sources and its potential role in enabling these joint observations.

\section{Evolution of White Dwarf Binaries} \label{sec:WDBevo}

The detection of WDB mergers and their possible final explosion is the focus of this work.
SN Ia 
can be used as standard candles, and their measurement is crucial in both astrophysics and cosmology. 
While the exact progenitor of SN Ia is still debated, the SD and DD scenarios are both considered plausible and leading contenders.  In fact, it is possible that some Type Ia's come from the SD scenario and some from DD scenario.
In the DD scenario, the supernova detonation originates from the merger of two WDs, whose orbit decays due to gravitational radiation. 
Detecting the GW signal of a WDB would allow us to localise and predict its merger time and position, paving the way to the observation of its EM emission and the eventual SN Ia.
An event with the combined detection of the GW and EM spectra is also referred to as a bright siren. 
Bright sirens are excellent cosmological probes, as they provide measurements of both the redshift and luminosity distance of the same objects.
Many cosmological parameters can be inferred 
by combining these two measurements of distance, such as the Hubble constant.
Currently, only one multi-messenger event involving GW has been observed: a binary neutron star coalescence detected with LIGO and Virgo in 2017, called GW170817, followed by a Gamma-ray burst measured with the Fermi Gamma-ray Space Telescope~\cite{abbott_multi-messenger_2017}. 
Atom-interferometers will provide a new opportunity to revolutionise multi-messenger astronomy, drastically expanding the sample of bright sirens by observing WDB mergers and predicting their luminous SN Ia. In fact, both MAGIS Space and AEDGE are expected to detect many WDBs and predict their coalescence, as highlighted in \autoref{fig:CharacteristicStrain}.
Beyond cosmology, multi-messenger observations of WDBs could also lead to breakthroughs in astrophysics, e.g., understanding the structure of WDs, the mass transfer dynamics in WDBs, and identifying the SN Ia progenitors.

In order to better understand the GW signature of WDBs to predict their coalescence and luminous event, in this section, we further discuss WDBs' properties, emphasising the ones relevant for this work. 
We start in \autoref{sec:WD} by giving an introduction to WDs and some specific properties of this compact object necessary for our analysis.
At first, we discuss the formation of WDs \autoref{sec:WDBformation}, then in \autoref{sec:WDBevolution} we present an overview of the evolution stages of WDBs, from formation to merger, while their population is discussed in \autoref{sec:WDBpopulation}.
Afterwards, we focus on the final stages. Firstly, in \autoref{sec:RLOF} we describe when the point mass approximation breaks down, then we model the mass transfer process in \autoref{sec:MT}. We conclude in \autoref{sec:SNformation} by explaining the possible outcomes after WDB coalescence.

\subsection{Introducing White Dwarfs}\label{sec:WD}
In this section, we briefly introduce white dwarfs as astrophysical objects and emphasise characteristics important to the present study~\cite{Carroll_Ostlie_2017}.

\begin{figure}[t]
\centering

\includegraphics[height=0.75\linewidth]{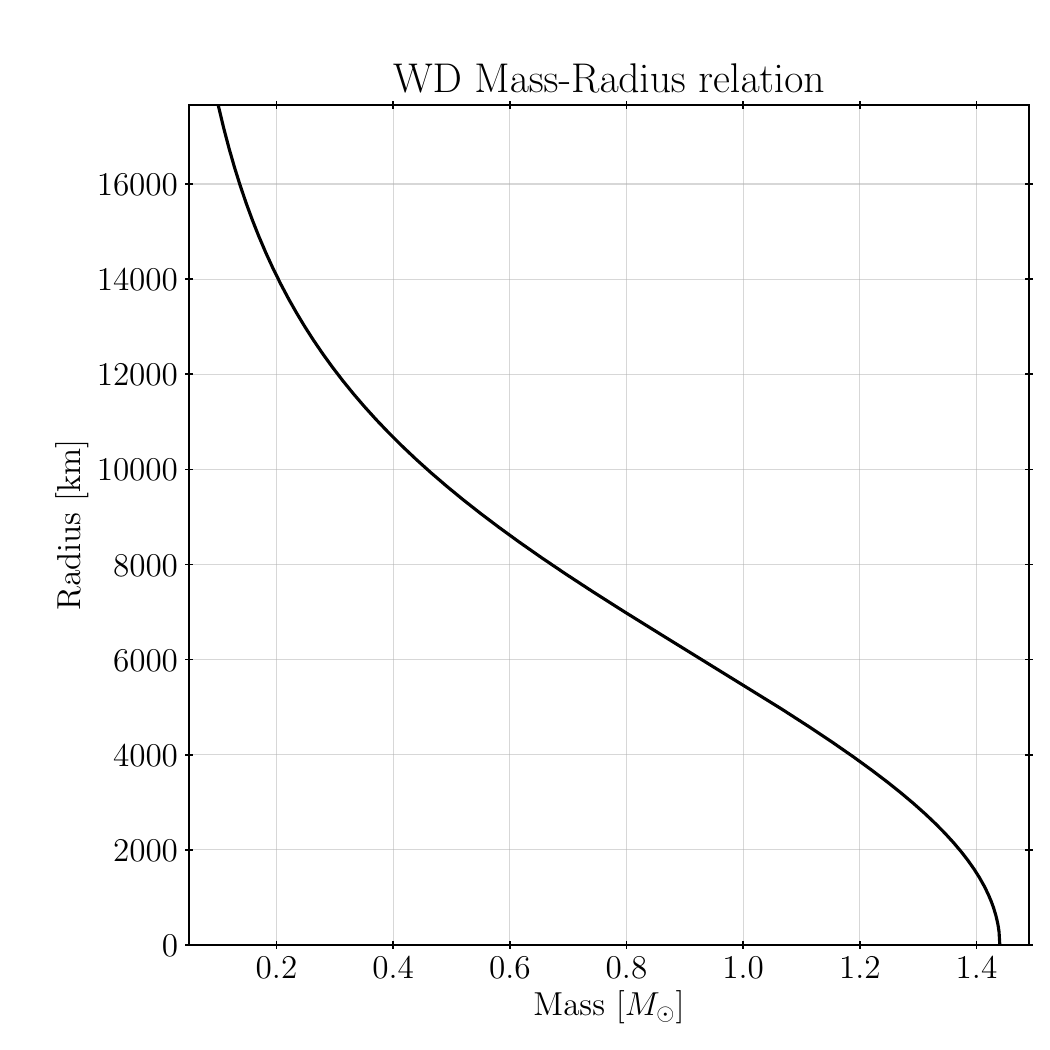}

\caption{Mass-radius relation of the white dwarfs, in a mass range between $0.1\,M_\odot$ and the Chandrasekhar limit, $1.44\,M_\odot$.
}
\label{fig:WDfig}
\end{figure}

White dwarfs (WDs) are extremely dense star remnants, stabilized against gravitational collapse by electron degeneracy pressure.
The equilibrium state is expected to be possible up to the Chandrasekhar mass $M_{Ch}\approx 1.44\, M_\odot$~\cite{Chandrasekhar_1931}, after which the degeneracy pressure cannot hold against gravity and the WD collapses. 
The WDs' equation of state is fairly well understood, and accordingly, their mass-radius relation can be computed with high precision (see e.g.~\cite{ambrosino2020white}). However, in numerous previous works, a good heuristic fitted version of WDs' mass-radius relation~\cite{Verbunt_1988} has been proven effective and is also used in this work~\cite{Marsh_2004,Kremer2017,toubiana2024interacting}. The mass-radius relation
\begin{align}\label{eq:WDmassradius}
    \frac{R}{R_\odot}=0.0114
    \left[
    \left(\frac{M}{M_{Ch}}\right)^{-2/3}
    -
    \left(\frac{M}{M_{Ch}}\right)^{2/3}
    \right]^{1/2}
    \left[
    1+3.5
    \left(\frac{M}{M_{p}}\right)^{-2/3}
    +
    \left(\frac{M}{M_{p}}\right)^{-1}
    \right]^{-2/3}~,
\end{align}
describes the non-trivial relation between the radius $R$ and mass $M$ of a WD, where $R_\odot$ is the solar radius, $M_{p}=0.00057\,M_\odot$ is a numerical constant, and $M_{Ch}$ the Chandrasekhar mass.
The WD mass–radius relation is illustrated in \autoref{fig:WDfig}. This relation is monotonically decreasing, meaning that more massive WDs are physically smaller in size.

WDs are primarily composed of carbon and oxygen (CO).
However, low-mass WDs often contain a higher fraction of helium (He), as their progenitor stars lack the mass required to ignite helium fusion~\cite{HeWD}.
At the opposite extreme, very massive WDs can exhibit significant neon (ONe) abundances, the result of progenitors hot enough to ignite carbon fusion before the WD stage~\cite{NeWD}.

\subsection{Introducing White Dwarf Binaries and their Evolution Phases}
The signal of interest for the atom-interferometer and  SN Ia explosions comes from the latest stages of the WDB lifetime.
Therefore, understanding their evolution up to this point is fundamental to predicting their gravitational wave signature and merger rates.

\subsubsection{Formation}\label{sec:WDBformation}
A white dwarf binary (WDB) forms when two main-sequence stars both evolve into white dwarfs. This transformation can occur through either a common-envelope phase or stable mass transfer~\cite{Nelemans_2001}, with the exact evolutionary pathway depending on the stars’ compositions. WDBs can be made up of helium–helium (He–He), carbon–oxygen–helium (CO–He), or carbon–oxygen–carbon–oxygen (CO–CO) pairs, with CO–CO systems being more massive. Extremely massive and rare WDs could have ONe composition, adding three possible combinations of binaries. Once formed, the two WDs are typically detached and stable, slowly spiraling toward each other as they lose energy through gravitational wave emission.

\subsubsection{Evolution}\label{sec:WDBevolution}
After their formation, the binary comprises two detached, stable WDs whose inspiral is driven by GW radiation. The frequency of the GWs, $f_{GW}$, evolves as
\cite{Maggiore:2007ulw},
\begin{equation} \label{eq:chirp}
    \frac{d f_{GW}}{dt}=\frac{96}{5} \pi^{8/3}\mathcal{M}_c^{5/3}f_{GW}^{11/3}~,
\end{equation}
where
\begin{equation}
    \mathcal{M}_c = \frac{(m_1m_2)^{3/5}}{(m_1+m_2)^{1/5}}~
\end{equation}
represents the chirp mass\footnote{The chirp mass at the source and the detector can change depending on the relative velocities. 
As we are analysing the incoming GW signal in this project, we consider $\mathcal{M}_c$ to be the chirp mass in the detector frame.}, which is related to the masses of the two WD components of the binary ($m_1$ and $m_2$), and determines the emission of gravitational energy, hence its ``chirp''. 
This phase can last $3$\,Gyr or longer, during which the binary gradually shrinks, being solely driven by gravity, thus radiating unperturbed GW radiation, because no tidal disruption or merger occurs yet~\cite{Nelemans_2001}. Since the system is slowly evolving, the WDs are detached and the orbit is almost circular, the quadrupole approximation remains valid~\cite{Yoshida_2021,Isern2003,Maselli_2020}, even when compared to numerical calculations~\cite{Postnov_2014}. So the WDs can be treated as point masses up until mass transfer occurs~\cite{Yang_2022}.
This treatment is possible mainly because the orbital separation remains large until the tidal disruption starts~\cite{Zou_2020}. As the WDs inspiral, they lose orbital momentum and the gravitational wave frequency can increase up to $1$\,Hz~\cite{Yang_2022,Maselli_2020,Zou_2020}. Due to the slow evolution of the inspiral process, the frequency signal is effectively monochromatic for the majority of the binary’s observable lifetime~\cite{toubiana2024interacting}.
The GW emission gradually increases its intensity during the inspiraling phase and reaches its peak power (maximum amplitude and frequency) right upon merger~\cite{Yang_2022}.

\subsubsection{Final phase}\label{sec:WDBfinal}
Following a very slow GW driven approach of the two WDs, 
they eventually become so close that other effects of the binary become relevant. 
This event is described by the Roche Lobe Overflow (RLOF), which, more specifically, represents the inability of the gravity of the less massive and less compact object to hold its mass together against the attraction of its companion. 
The matter on the outer layer of the donor starts flowing towards the more compact star. 
More details on RLOF are reported in \autoref{sec:RLOF}, as it represents a fundamental turning point in WDB evolution.
The mass transfer is expected to begin slowly and intensify smoothly with the WDs getting drawn together again due to gravitational radiation. 
However, from RLOF on, the binary evolution becomes more complicated as a result of the interplay between multiple factors. 
In fact, the exchange of material might have different outcomes depending on the efficiency and the size of the objects, which can either favour or disfavour the collapse. 
In this paper, we call binaries that are favoured to merge `unstable', while the ones that tend to survive for a longer time are called `stable'.
Many effects during mass transfer can only be approximated, while others are neglected because of their minor influence and excessive complexity. More information is found in \autoref{sec:MT}.

Different combinations of WD masses, composition and dynamics lead to variable WDB lifetimes and various final events, which remain difficult to predict.
The system may end in a merger, detonation, collapse to a more compact object, outspiral again, or settle into a stable AM Canum Venaticorum star (AM CVn) system, an ultra-compact binary in which a WD accretes helium-rich material from a low-mass donor.
If a WDB mass transfer approaches the critical mass transfer threshold, the lighter WD will be torn apart shortly after, and a merger is to be expected~\cite{Dan2012, Pakmor_2012, Sato_2016}. 
Several possibilities have been proposed for the outcomes, and the most common are SN Ia, collapse into a neutron star, formation of a larger WD, or minor other Novae~\cite{Yungelson2016, Shen_2015, Marsh_2011}. 
The vast spectrum of possible aftermath is further examined in \autoref{sec:SNformation}.

\subsection{White Dwarf Binary population}\label{sec:WDBpopulation}

Additional information regarding WDs, which are relevant for this project, includes their distribution both in space and mass. More specifically, we need these details for WDBs and not for the singular objects.
Various projects tried to estimate the binary distribution~\cite{Nelemans_2001_galax,Marsh_2011,Badenes:2012ak,Toonen_2012,Rebassa_Mansergas_2018,  Maoz:2018epf,Breivik_2020,DWDmasses}, but a consensus has not been reached. While the singular WD mass function has been measured and well-studied
\cite{WDmass07,WDmass13,WDmass16,WDmass231,WDmass232,WDmass24}, a precise observation of the binary components is still missing. Distinguishing between individual WDs and binary systems is extremely hard due to their faint light emission. 
Additionally, a focused survey on WDBs has not happened yet. Another approach to estimate the WDB population is based on SN Ia. Recent surveys tried to compute the number and rates of SN Ia events~\cite{Frohmaier2019,Zwicky2020,Sharon2021}. Since we may expect at least some of these detonations to be ignited by WDBs, their volumetric rate can be used to infer the spatial distribution of WDBs, as is discussed in \autoref{sec:mergerrate}. The measured SN Ia rate is $2.43\times 10^{-5}\ \text{yr}^{-1}\text{Mpc}^{-3}$ (\ref{eq:SNIarate}).

Additionally, we also use a population synthesis model to estimate the mass distribution of WDBs. We recover the binaries' mass function from simulated populations based on star formation models and tested with the limited WDB observations available~\cite{Seppe24,Hofman24}. This theoretical simulation is based on the code \texttt{White\_Dwarf\_AGWB} \href{https://github.com/SeppeStaelens/White_Dwarf_AGWB}{\faGithub} \cite{StaelensWhiteDwarfAGWB2024}, which uses the code \texttt{SeBa}~\cite{ZwartVerbunt1996,Nelemans_2001,Toonen_2012}, that simulates stars and binary evolution.
The code models the amount and properties of WDBs starting from a homogeneously distributed stellar mass with a characteristic mass function~\cite{Krupa01}. Uncertainties on this model arise from the star formation rate densities used, which vary with their metallicities, parametrised with $Z$, and the capacity of binaries to transfer mass and angular momentum. 
This simulation models the WDB formation in two phases parametrized by either a common envelope parameter $\alpha$ or an angular momentum balance parameter $\gamma$.
Articles~\cite{Seppe24,Hofman24} provide the results for 4 different models, $\alpha\alpha$ and $\gamma\alpha$ with $\alpha=1$ or $\alpha=4$~\cite{Nelemans_2001,Toonen_2012,NelemansTout2005}, and 3 star formation rates based on 6 metallicities, from $Z=0.0001$ to $Z=0.03$~\cite{Chruslinska2019,Chruslinska2020,Chruslinska2021,madau_cosmic_2014}.
The model that best fits the observation has $\gamma=1.75$ in the first phase and $\alpha=4$ in the second~\cite{Hofman24}. 
Furthermore, the fiducial star formation rate history is taken from~\cite{Chruslinska2019}. In the present Universe, the metallicity $Z=0.02$ is estimated to dominate.  

More details on the implementation of the various models and metallicities in our project to estimate WDB's merger rate are discussed in  \autoref{sec:mergerrate}.

\subsection{Roche Lobe Overflow}\label{sec:RLOF}
The components of a binary can be considered point masses for most of their lifetime, as their evolution is primarily driven by gravity and angular momentum. In the late stages of the dual system, as in the case of WDB, the point-mass approximation loses its validity as the two elements start exchanging mass. This event is of major importance in describing the final phase. The compact objects' sizes, $R_1$ and $R_2$, and masses, $M_1$ and $M_2$\footnote{By definition in this work we always have $M_1>M_2$, see \autoref{tab:parameters}.}, play a fundamental role in this analysis, together with the Roche potential $\Phi$, which approximates the total potential energy of the system~\cite{Postnov_2014}. In a Cartesian frame of reference rotating with the binary, with the two objects at $\Vec{x}_1=(0,0,0)$ and $\Vec{x}_2=(a,0,0)$, separated by a distance\footnote{It is common to use $a$ as the major semi-axis when discussing binaries. However, due to the late stages of the binaries of interest and the GW emission evolution described above, we consider the binary to have circular orbits. In this specific case, the eccentricity is null $e=0$ and the major semi-axis of the binary is equal to the distance between the two compact objects.} $a$, the Roche potential in an arbitrary point $\Vec{x}=(x,y,z)$ is the following,
\begin{equation}\label{eq:RochePotential}
    \Phi= -\frac{GM_1}{\sqrt{x^2+y^2+z^2}}-\frac{GM_2}{\sqrt{(x-a)^2+y^2+z^2}}
    -\frac{1}{2}\Omega_{\rm orb}^2\left[(x-\mu a)^2 + y^2\right]~,
\end{equation}
where $\Omega_{\rm orb}$ is the angular frequency of the rotation and $\mu=\frac{M_1}{M_1+M_2}$.
The tightest Roche equipotential line encircling both masses forms the Roche lobes. 
When one companion reaches this threshold, its outer layer is free to orbit around both compact objects, and the point mass approximation loses its validity. 
Thereafter, the mass transfer is dependent on the extent the object crosses the Roche lobe, also called Roche Lobe Overflow (RLOF). 

\begin{figure}[t]
    \centering
    \includegraphics[width=\textwidth]{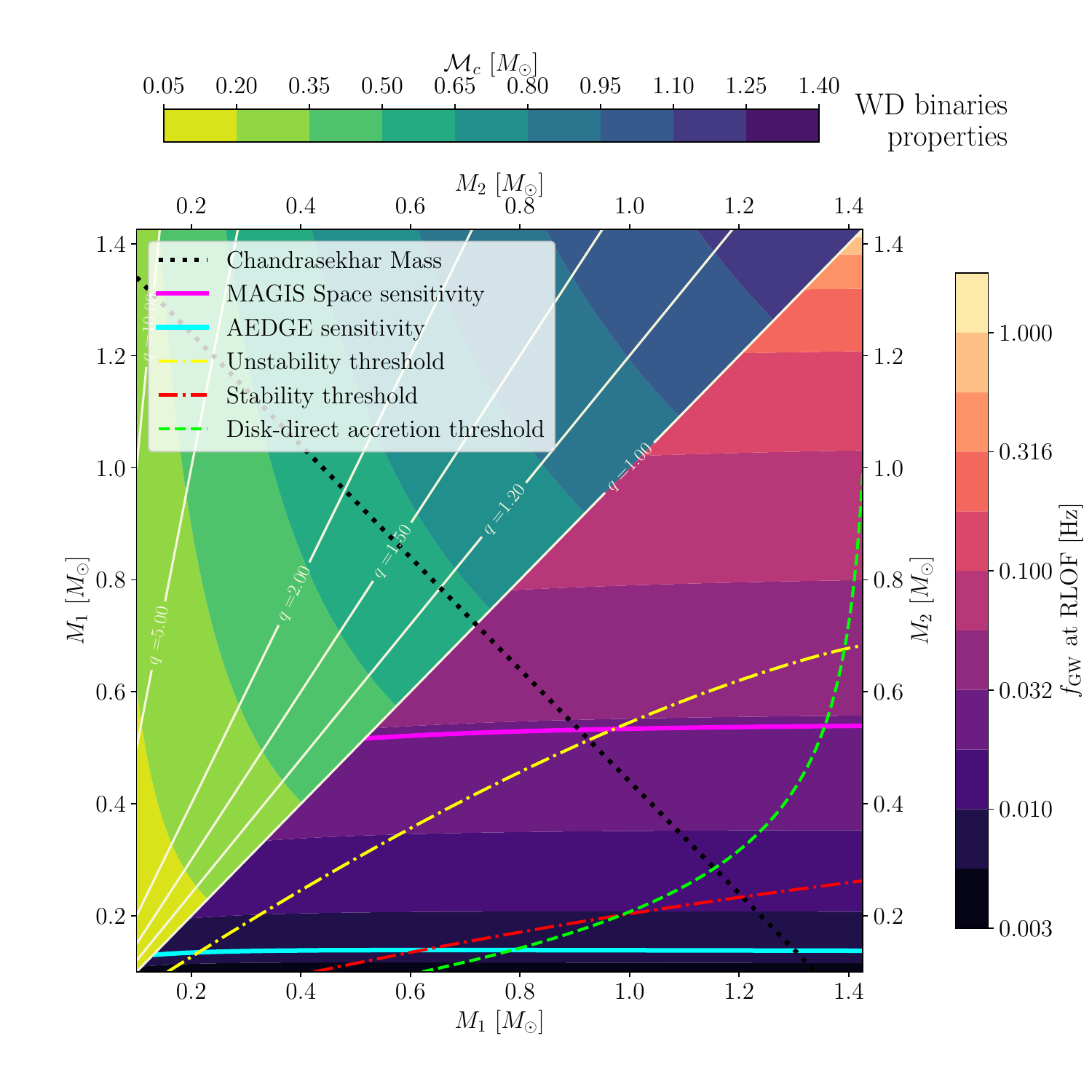}
    \caption{WDB properties. On the top left part of the plot, the relation between individual masses of the binaries $M_1$ and $M_2$, chirp mass $\mathcal{M}_c$ and mass ratio $q$ is displayed. On the lower right triangle, the frequency at which each binary's mass combination hits the Roche Lobe Overflow is shown. In addition, a few extra pieces of information are displayed: the Chandrasekar mass (dotted black line), the lightest WDBs that can be observed by MAGIS Space and AEDGE instruments based on the frequencies at RLOF (respectively, solid pink and cyan lines), and some thresholds 
    related to the stability of the binary~\cite{Marsh_2004}. 
    More specifically, the plot includes the lowest and highest limits for which WDBs can stabilise, depending on the binaries' properties, and avoid merger (dashdotted red and yellow lines), and the highest masses for which an accretion disk can form (dashed green line).}
    \label{fig:WDB_RLOF}
\end{figure}

The exact moment at which mass transfer begins cannot be computed analytically from equation~\eqref{eq:RochePotential}. 
A solution for the initial condition of the RLOF, precise up to $1\,\%$, was provided by Eggleton~\cite{Eggleton_1983} and has been proven successful in many models and applications~\cite{Podsiadlowski_2014}. 
Eggleton's formula reports the relation between the radius of the Roche lobe $R_{L}$ around $M'$, assumed as spherical\footnote{Despite the non-spherical shape of the Roche lobe, this solution approximates the lobe to a sphere because of the shape of the donor and the effectiveness of the result.}, and the distance between the centres of the two compact objects $a$.
\begin{equation}\label{eq:RocheLobe}
    \frac{R_{L}(a,q_L)}{a}=\frac{0.49 }{0.6+q_{L}^{2/3}\ln\left(1+q_{L}^{-1/3}\right)}~,
    \qquad
    0<q_{L}<\infty~,
\end{equation}
where $q_{L}=\frac{M''}{M'}$ is the mass ratio of the two bodies.
The distance between the two compact objects at which mass transfer starts, $a_{RLOF}$, 
is found by setting the Roche lobe radius equal to the donor's size $R_2$,
which for WDBs is always the lighter companion $q_{L\,2}=\frac{M_1}{M_2}=\frac{1}{q}$,
\begin{equation}\label{eq:RLOFcond}
    R_{L}(a_{RLOF},q_{L\,2})=R_2~.
\end{equation}
The same equation~\eqref{eq:RocheLobe} can also be used to compute the Roche lobe radius around the other WD, by using the inverse mass ratio $q_{L\,1}=\frac{M_{2}}{M_{1}}=q$, although mass transfer always starts at $R_2$ due to the monotonic mass-radius relation of these objects~\eqref{eq:WDmassradius}.
The intensity of mass transfer is characterised by the extent of RLOF, hence the difference between the radius computed above and the actual radius of the object, and is defined as follows:
\begin{equation}
    \Delta(a)=R_{L}(a,q_{L\,2})-R_{2}~.
\end{equation}

It is also worth mentioning the implications of RLOF on the astrophysical phenomenon of interest: white dwarf binaries. The Roche lobes are always wider for the more massive binary components, due to their stronger gravity. For other compact celestial objects, it might be complex to understand which companion is the donor because of their non-trivial mass-radius relation. Nevertheless, identifying the donor is straightforward for WDBs. Indeed, WDs have a monotonically decreasing size with increasing mass, equation~\eqref{eq:WDmassradius}, as shown in \autoref{fig:WDfig} ($M_1>M_2$ means $R_1<R_2$). The less massive object of a WDB has both a smaller Roche lobe and a wider radius, hence it will overflow sooner than its companion, becoming the donor in the mass transfer process.

The frequency at which RLOF starts can be computed using equation (\ref{eq:RocheLobe}) and Kepler's law, as the underlying assumption that the binary has a circular orbit, due to the long period of GW-driven evolution, stands.
\begin{equation}\label{eq:RLOFfreq}
    f_{GW,RLOF}=2f_{RLOF}=\sqrt{G\frac{M_1+M_2}{a_{RLOF}^3\pi^2}}~,
\end{equation}
where $f_{RLOF}$ is the orbital frequency of the binary 
and $f_{GW,RLOF}$ is the frequency of the emitted GW.
\autoref{fig:WDB_RLOF} shows the frequencies of WDBs when they reach RLOF, depending on the mass of the two WDs.

\subsection{Mass transfer}\label{sec:MT}
Once two compact objects come sufficiently close to fill their Roche lobes, mass transfer between them begins.
As the separation decreases, primarily due to gravitational wave emission, the intensity of this mass exchange gradually increases.
Ultimately, all binaries are expected to merge or be disrupted, though the timescale for this outcome can vary greatly depending on the system's characteristics.
According to the literature, a binary will collapse rapidly if the donor exceeds a critical mass transfer rate of $0.01\,M_\odot {\rm ~yr}^{-1}$~\cite{Marsh_2004,Kremer_2015,Kremer2017,toubiana2024interacting}. In this paper, we frequently refer to this moment as the merger or coalescence time, since it can be quantified and precedes the final event by a very short interval. 
We begin by outlining the key competing factors that determine the stability of a binary system. These factors influence whether the system becomes unstable and collapses rapidly, or remains stable, allowing for a longer inspiral until merger.
\begin{itemize}
    \item \textit{Gravitational Radiation}: Gravitational wave radiation shrinks the orbit of the binary. Thus, the orbital frequency increases. Overall, the system loses energy, and consequently, its angular momentum also decreases. 
    GWs drive the binary's evolution before RLOF and also shortly after, until mass transfer takes over.
    \item \textit{Mass transfer}: After RLOF, mass transfer slowly increases and soon drives the dynamics of the dual system. 
    In contrast to GWs, the mass exchange can make the binary unstable or stable. 
    \begin{itemize}
        \item \textit{Unstable}: Due to the inverse mass-radius relation for WDs (see \autoref{fig:WDfig}), the donor grows in size by losing mass. 
        As the overlap of their Roche Lobes increases, so does the matter loss from the donor, leading to even greater instability as the system approaches the critical mass transfer rate. 
        However, this could subsequently stabilise the binary by boosting the next effect.
        \item \textit{Stable}: As the donor loses mass, the mass distribution within the system shifts.
        To conserve angular momentum, the two WDs are driven apart, slowing down mass transfer and making the system more stable. 
        The extent of this effect's influence depends on the efficiency of mass transfer and the accretion onto the more massive WD.
    \end{itemize}
    \item \textit{Spin-orbit coupling}: 
    The spin of a WD can exert a torque on the system through non-trivial tidal and magnetic interactions with the orbit. 
    As a consequence, WDs generally tend to synchronise their rotation with the binary orbit. 
    This effect is relatively weak and normally relevant only over long timescales. 
    Until RLOF, the contributions of GWs and spin–orbit coupling gradually reach equilibrium, leading to synchronisation of the compact objects. 
    However, once mass transfer begins, the orbital frequency can vary more rapidly, causing the individual WDs to lose synchrony with the binary motion.         
    At this point, the impact of spin–orbit coupling counteracts this desynchronisation. Nevertheless, its effect remains secondary as it does not grow as rapidly as its competitor mechanisms (GWs and mass transfer). 
\end{itemize}
It follows that the angular momentum of the binaries and the efficiency of mass transfer play an important role in the stability of the system. 
Therefore, the dynamics of WDBs, both prior to and following RLOF, can be analysed by considering the conservation of total angular momentum. This total consists of the orbital angular momentum of the binary, $J_{orb}$, and the spins of the WDs, $J_1$ and $J_2$. 
In addition, the binary framework can lose angular momentum only by radiating GWs $\dot{J}_{GW}$ or losing matter $\dot{J}_{loss}$~\cite{Marsh_2004,toubiana2024interacting}
\begin{equation}
    \dot{J}_{orb}+\dot{J}_1+\dot{J}_2=-\dot{J}_{GW}-\dot{J}_{loss}~.
\end{equation}
Since our goal is to estimate the post-RLOF merger timescale, we follow the simplifying assumptions of~\cite{Marsh_2004}, neglecting secondary effects and detailed parameters of the system’s evolution. 
First, we neglect both the angular momentum evolution of the spins $\dot{J}_{2}=0$ as well as the one from total mass loss $\dot{J}_{loss}=0$, due to their expected small contributions. Secondly, we assume that all matter leaving the donor accretes on the more massive WD $\dot{M}_2=-\dot{M}_1<0$.
Additionally, the mass loss rate of the donor is dependent on the RLOF $\dot{M}_2\propto {\Delta}^3$~\cite{Marsh_2004,Webbink1984}, where the overflow $\Delta$ was discussed in the previous section, see equation~\eqref{eq:RLOFcond}.
Other references present more detailed computations~\cite{Kremer_2015,Kremer2017,toubiana2024interacting,Gokhale_2007}.
The evolution of the WDB orbit after RLOF due to mass transfer can be simplified as follows~\cite{Marsh_2004}:
\begin{equation}
\label{eq:masstransferorbit}
    \frac{\dot{a}}{2a}= -\frac{32}{5}\frac{G^3}{c^5}\frac{M_1M_2(M_1+M_2)}{a^4} 
    -\left(1-1/q-\sqrt{r_h(1+1/q)}\right)\frac{\dot{M}_2}{M_2}
    +\frac{k M_1 R_1^2}{\tau_S J_{orb}} \omega~,
\end{equation}
where
\begin{itemize}[label=\tiny$\bullet$, itemsep=1pt, parsep=0pt]
    \item $a$ is the distance between the two WDs;
    \item $k$ describes the moment of inertia of the donor (fitted to $k=0.1939(1.44885-M_1)^{0.1917}$~\cite{Marsh_2004});
    \item $R_1$ represents the radius of the accretor;
    \item $\tau_S$ is the tidal synchronization timescale; 
    \item $J_{orb} = \sqrt{\frac{Ga}{M_1+M_2}}M_1M_2$ introduces the angular momentum of the binary;
    \item $\omega=\omega_{WD_1}-\omega_{WDB}$ is the difference in angular frequency between the accretor's spin and the orbit of the binary \footnote{Following~\cite{Marsh_2004} $\omega$ is expected to be positive};
    \item 
    $r_h a$ is the widest radius at which accreting matter could orbit the more massive WD, and it has been fitted to $r_h=0.0883+0.04858\log(q)+0.11489\log(q)^2-0.020475\log(q)^3$~\cite{Verbunt_1988}. The term in equation~\eqref{eq:masstransferorbit} including $r_h$ describes the angular momentum transfer between the two WDs. 
\end{itemize}

\begin{figure}[t]
\centering

\includegraphics[width=0.7\linewidth]{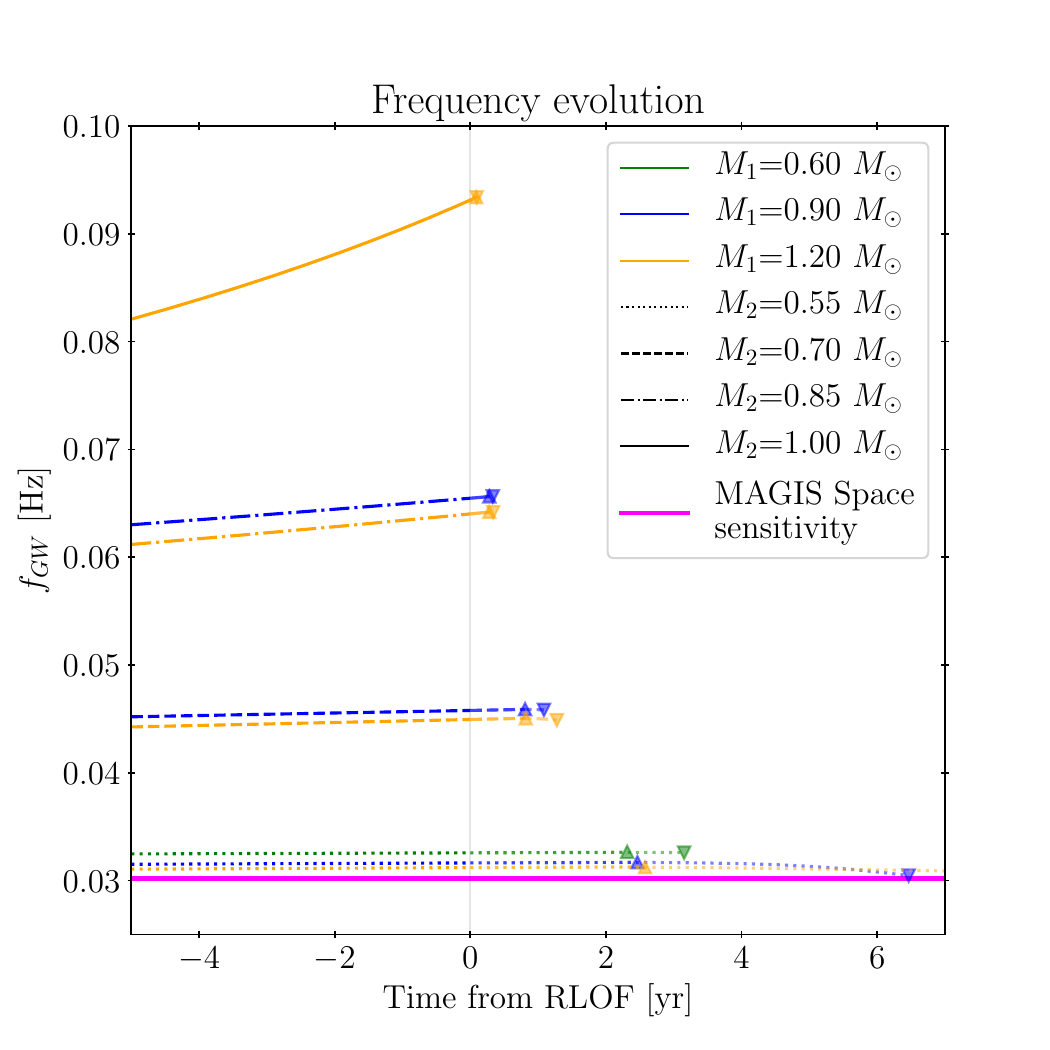}
\caption{
Frequency evolution of various characteristic WDBs, before and after RLOF, which is represented by the vertical line. Before RLOF, or on the left-hand side of the grey line, the WDB evolution is totally GW driven~\eqref{eq:chirp}, hence the curve shows the chirp of the system. After RLOF, or on the right-hand side of the vertical line at time $0$, the frequency evolution follows equation~\eqref{eq:masstransferorbit} including all the competing factors. For each WDB, after RLOF, the lines are split into two, representing the maximally stable and unstable frequency evolutions. The two regimes can be recognised by their disruption time, which is displayed with a triangle pointing upwards in the stable case, and with one pointing downwards in the unstable scenario.
The colours represent the mass of the most massive WD in the binary, while the line style indicates the mass of the lighter companion. Only systems with $M_1>M_2$ are shown. }
\label{fig:Frequencies}
\end{figure}

The three terms on the right hand side of equation~\eqref{eq:masstransferorbit} describe, in order, the competing factors listed above:
from the left, shrinking of the orbit with GW radiation, mass transfer effects, which could make the system more or less stable, and spin-orbit coupling.
Due to the significant uncertainties surrounding the properties of WDBs, the influence of the last two effects on the system's evolution cannot be predicted.
To approach this challenge, we examine two extreme scenarios (representing the most stable and the most unstable cases) to understand the possible evolutionary timescales of the binary. Realistic cases are expected to fall between these two limiting cases.
\begin{enumerate}[label=\roman*.]
    \item \underline{Stable case}: in this scenario, the spin-orbit coupling of the accretor plays a relevant role, exerting a torque on the system and stabilising it.
    From equation~\eqref{eq:masstransferorbit}, it can be noticed that a WDB is most stable when the momentum transfer term ($\propto r_h$) and the spin-orbit coupling term ($\propto \omega$) cancel each other out.
    In this particular scenario, all the angular momentum transferred from the donor is fed back into the orbit due to the spin-orbit coupling.
    \item \underline{Unstable case}: we consider a direct impact accretion with no dissipative torque. Spin-orbit coupling provides no stabilising effect, as we assume the system remains synchronised ($\omega=0$). Furthermore, mass transfer is considered perfectly efficient, conserving both mass and angular momentum. 
\end{enumerate}
The mass-radius relation is well understood, thus when evaluating equation~\eqref{eq:masstransferorbit} we use a numerical implementation of equation~\eqref{eq:WDmassradius}.

\autoref{fig:Frequencies} shows the evolution of a series of WDBs before and after RLOF obtained by numerically integrating equation~\eqref{eq:masstransferorbit} and highlights the evolution for the most stable and unstable scenarios. An actual binary is expected to merge in a timespan between these two cases.
One can observe that mass transfer and binary stability appear to play a more significant role in relatively lighter WDB systems. 
In contrast, more massive binaries experience stronger gravitational effects and typically merge rapidly after RLOF, driven by GW evolution.
It must also be noted that article~\cite{Marsh_2004} provides analytical thresholds for instability and stability, which are included in \autoref{fig:WDB_RLOF}. Additionally, an accretion disk may form if the transferred mass does not directly impact the accretor. As shown in \autoref{fig:WDB_RLOF}, the range of binaries affected by this phenomenon is not within the interest of our work; hence, the accretion disk is not considered.
Lastly, it must be mentioned that the properties of WDs in binaries that define their dynamics (e.g. spin, accretion, \dots) are variable.
Therefore, the period before the inevitable coalescence can span a broad range of timescales depending on the binary stability.
MAGIS Space will represent an incredibly useful instrument for the study of the late-stage evolution of WDB by measuring the dynamics after RLOF through its GW signatures.

\subsection{WDB outcome}\label{sec:SNformation}
The latest stages of the binary between when mass transfer becomes critical ($-\dot{M}_2>0.01\,M_\odot\,{\rm yr}^{-1}$) and the coalescence are of great interest, as well as their final effects and outcome. Some studies have tried to predict this with hydrodynamical simulations~\cite{Dan2012,Pakmor_2012,Sato_2016}. Several different events remain possible for describing the end of these binaries, depending on the stability of mass transfer, their masses and their compositions, and could or could not lead to Type Ia supernovae~\cite{Shen_2015,Yungelson2016,Sato_2016}. Some of the possible products are the formation of a more massive WD, the collapse of a neutron star, SN Ia, but also many other fainter thermonuclear explosions.

MAGIS will measure the GW signature of the latest stages of any WDB loud enough to be detected, predicting their coalescence and shedding light on the different types of runaways and their conditions, for SN Ia as well as other outcomes. 
In this paper, we focus mainly on the possible progenitors of SN Ia, since these events are of great interest to cosmology. 
SN Ia are very energetic explosions expected to be generated from a runaway fusion of WD carbon-oxygen cores, which completely disrupts the star. 
These detonations are peculiar because of their unique spectral features, which present no hydrogen or helium spectral lines, making them easily recognisable.
Additionally,  
they are considered standard candles as they are particularly suitable for measuring distances, due to their very consistent brightness curve and duration. 
The SN Ia afterglow lasts for weeks, however, the earlier it is observed, the more information can be obtained.  Having the GW signal would allow prediction of when mergers would occur, allowing observations to be made before and during the merger/SN Ia event, thus improving the information available.
In the DD scenario, the conditions to produce SN Ia can arise in multiple situations. For example, one option is that, due to accretion, the internal pressure of the accretor increases, together with its mass, triggering the detonation. Another option is that the SN Ia is triggered by the shock of a first explosion from helium that has accumulated in an outer layer. This latter prospect is called the dynamically driven double-degenerate double-detonation scenario~\cite{Kosakowski2022}. The possibilities that lead to SN Ia are several and complex, making it hard to anticipate the combinations of WDB masses which act as their progenitors.
For this reason, following the article~\cite{Shen_2015}, we selected two regions in the WDB masses space $M_1-M_2$ for producing the supernovae of interest, which depict two distinct and extreme scenarios:
\begin{enumerate}[label=\roman*.]
    \item \underline{Conservative SN Ia scenario}: assume the only WD mass combinations that are expected to directly produce SN Ia explosions are:
    \begin{align}\label{eq:SNconservative}
        0.8\,M_\odot\leq&M_1\leq 1.1\,M_\odot~, &
        0.1\,M_\odot\leq&M_2\leq 1.1\,M_\odot~, &
        M_1+M_2&\geq 1.2\,M_\odot~.
    \end{align}
    \item  \underline{Optimistic SN Ia scenario}: assume a broader spectrum of  WDB mass combinations can generate SN Ia, either directly or through secondary effects:   
    \begin{align}\label{eq:SNoptimistic}
        0.5\,M_\odot\leq&M_1\leq 1.44\,M_\odot~, &
        0.1\,M_\odot\leq&M_2\leq 1.1\,M_\odot~.
    \end{align}
\end{enumerate}
These regions were chosen as they represent the two far ends of WDB mass regions that go SN Ia, hence it is reasonable to expect that the real scenario will fall in between these two extreme cases.
An illustrative description of these mass regions is shown in \autoref{fig:WDBweighted}.

Some articles~\cite{Shen_2015,Yungelson2016,Breivik_2020} have different claims, for example, that SN Ia is even rarer than our conservative scenario, and the combinations of WDs that lead to this outcome are only made of the more massive ones. Other references studied the dynamics of specific WDBs~\cite{Pakmor:2013wia, Boos:2021acp, Pakmor:2021, Shen:2021bwu, Pakmor:2022lwn, Kosakowski2022}. Nevertheless, our estimate is expected to be reliable, since the highest contribution to the detectable WDB that goes SN Ia comes from the more massive and louder WDB. The mass dependence contribution to the detectable merger rate with MAGIS Space is discussed in \autoref{app:contributions}, and displayed in \autoref{fig:Mergerratescontributions}.
One could even argue that our selection is more conservative. 
In fact, a more restricted WDB mass region that includes only the more massive binaries from the conservative scenario would lead to more detectable GW signatures before SN Ia when considering the same SN Ia rate and DD scenario contribution to it.
\\[1\baselineskip]
In this section, we have introduced all the elements relative to WDs and their binary systems required to further explain our study on their GW and EM signatures and possible future detections.
In order to streamline our analysis and be conservative, we studied limit regimes with simplified dynamics models.
In particular, we defined two extreme cases in terms of dynamics after RLOF, the stable and unstable cases in \autoref{sec:MT}, and two in terms of SN Ia production from WDB mergers, the conservative and the optimistic SN Ia scenarios in the last segment.
For more information regarding the limitations of these simplifications, refer the reader to \autoref{app:corrections}.
These different scenarios will give us a picture of the range of possible observations, driving the uncertainty of the final results.

\section{GW signals from White Dwarf Binaries} \label{sec:GW}

The gravitational wave signal of WDBs, more specifically their merger, is one of the most interesting and novel astrophysical events that MAGIS Space is expected to observe.
White Dwarfs are less compact than neutron stars and black holes and thus merge at lower frequencies.  Hence ground-based laser interferometers (both LVK and new generation CE and ET) cannot see such mergers. Additionally LISA is not expected to be sensitive enough at the frequencies at which mergers of WDBs could be detected $3$\,mHz-$1$\,Hz (see \autoref{fig:CharacteristicStrain}) to actually see merger events, though it will see many WDBs long before merger.
One feature of great interest related to these celestial objects, both in astrophysics and cosmology, is the possibility of measuring two independent and complementary indicators of distance from the same event, redshift and luminosity distance.
This opportunity arises from the fact that WDBs are among the most likely progenitors of SN Ia. 
The GW signature from WDB can be used to study various properties of WDs and inspect the latest dynamics of their binary systems, but this signal can only provide a value for the luminosity distance.
An EM counterpart is necessary to determine the redshift.
To obtain a dual distance measurement, telescopes must be alerted in time regarding a potential SN Ia explosion and be able to target the source precisely.
Thus, the parameter reconstruction from the GW observations of MAGIS Space has to be sufficiently precise to localise and predict the time of coalescence of WDB.

In this section, we describe the GW signal from inspiraling WDBs before RLOF and forecast the accuracy on the different GW parameters that can be achieved with MAGIS Space.
We start by introducing the model used to describe the GW signals in \autoref{sec:parameters}.
In the following \autoref{sec:GWsignal}, we study the parameter characterisation of WDBs at a representative distance. 
This analysis provides us with a prediction on the localisation of the source as well as the loudness of the signal. The same method is then used in sections \ref{sec:Mergersignal} and \ref{sec:mergerrate} to estimate, respectively, how well we can anticipate SN Ia explosions and the observable WDB merger rate.

\subsection{GW signal modelling and Fisher analysis} \label{sec:parameters} 

We start by introducing the WDB GW signal and quantifying the information that these new instruments could provide us with.
We are interested in the GW signature of the final phase of the inspirals of WDBs with masses $0.1-1.5\,M_\odot$, which translates to $3$\,mHz-$1$\,Hz. 
The dynamics of such systems up until RLOF are well-described by the leading terms in the post-Newtonian (PN) expansion of General Relativity (for higher order corrections, see for example~\cite{Maggiore:2007ulw}).
The frequency evolution of the GW signal is described by equation~\eqref{eq:chirp}, and reported here again,
\begin{equation}\label{eq:chirp2}
    \frac{d f_{GW}}{dt}=\frac{96}{5} \pi^{8/3}\mathcal{M}_c^{5/3}f_{GW}^{11/3}~,
\end{equation}
where $\mathcal{M}_c$ is the chirp mass. The frequency of the GW signal,  $f_{GW}(t)$, at leading order, can be obtained by integrating equation~\eqref{eq:chirp2}.
The GW strain signal $h$ of the binary has two polarisations: $+$-polarisation and $\times$-polarisation. We study GWs in the transverse-traceless (TT) gauge, a particularly convenient framework for decomposing the two polarisation modes
\begin{equation}
    h_{\mu\nu}(t)=h_+(t)e_{\mu\nu}^+ + h_\times(t)e_{\mu\nu}^\times~,
\end{equation}
where $e^+$ and $e^\times$ are the polarisation tensors in TT gauge. 
To understand the interaction between the interferometer and GW, the so-called antenna functions $F(t)$ are introduced. This element couples the GW polarisations to the instrument direction $\hat{\textbf{l}}$, which in our case is made of a single baseline, describing its response to the GW signal,
\begin{equation}
    F_{+,\times}(t)=\sum_{i,j}\hat{l}_i(t) e^{+,\times}_{ij} \hat{l}_j(t)~.
\end{equation}
Consequently, the GW strain signal measured by the atom-interferometer can be computed,
\begin{equation}
    h(t)=h_+(t)F_+(t) + h_\times(t)F_\times(t)~.
\end{equation}
The antenna functions $F_{+,\times}(t)$ depend on the position and direction of the interferometer $\hat{\textbf{l}}$, which evolves with time. These functions also relate to the position and orientation of the GW source in the sky, which, on the other hand, can be considered to be fixed in time. The explicit parameters are the right ascension $\alpha$ and the declination $\delta$, which describe the localisation of the binary in the sky\footnote{In this project, we use a heliocentric equatorial coordinate system.}.
In addition to that, the polarisation angle $\psi$, which defines the orientation of the polarisation axes on the plane of the sky, is also a variable in the antenna function.

In the TT gauge, the GW strain in these two polarisations $h_+$ and $h_\times$ has a simple form at leading order,
\begin{align}\label{eq:hplus}
    h_+(t)&=\frac{ 2 \mathcal{M}_c^{5/3} \pi^{2/3}f_{GW}(t)^{2/3} }{d_L}
    \left[1+\cos\iota ^2\right]\cos\Phi(t)~,
    \\\label{eq:hcross}
    h_\times(t)&=\frac{ 4 \mathcal{M}_c^{5/3} \pi^{2/3}f_{GW}(t)^{2/3} }{d_L}
    \cos\iota\sin\Phi(t)~,
\end{align}
where $d_L$ is the luminosity distance of the source and $\iota$ represents the inclination angle between the line of sight and the angular momentum of the compact binary of interest.
$\Phi$ represents the phase accumulated by the binary from a time of reference $t_0$ up until $t$,
\begin{equation}\label{eq:phase}
    \Phi(t)=\int_{t_0}^t 2\pi f_{GW}(t') dt' + \Phi_0~,
\end{equation}
where $\Phi_0=\Phi(t_0)$\footnote{As $t_0$ is a free parameter, for simplicity, this is often set to be equal to $t_c$. $t_c$ is the time of coalescence if one would consider the compact objects to be point masses. In other words, $f_{GW}$ at $t_c$ diverges. In fact, this time reference is used to compute the GW frequency from equation~\eqref{eq:chirp2},
\begin{equation}
    f_{GW}=\int_{t}^{t_c} \frac{d f_{GW}}{dt}~,
\end{equation}
at an arbitrary time $t$. An alternative and convenient time variable often used in GW physics is $\tau=t-t_c$.}.

A summary of the parameters for the GW signal detection is shown in \autoref{tab:parameters}.

\begin{table}[t]
    \centering
    \begin{tabular}{c|l}
    Symbol & Parameter \\
    \hline
        $\mathcal{M}_c$ & Chirp mass \\
        $q=M_1/M_2$ & Mass ratio ($M_1>M_2$)\\
        $d_L$ & Luminosity distance \\
        $\Phi_0$ & Binary phase at $t_0$ \\
        $t_0$ & Time of reference \\
        $\iota$ & Inclination angle \\
        $\psi$ & Polarization angle \\
        $\alpha$ & Right ascension \\
        $\delta$ & Declination \\
    \end{tabular}
    \caption{The nine parameters that describe the waveform of a compact binary.}
    \label{tab:parameters}
\end{table}

To estimate the capabilities of future atom-interferometer GW detectors to measure the properties of a WDB from the GW signal, we use a standard Fisher Matrix study. 
This analysis is implemented in a tool similar to  \texttt{AIMforGW}  \href{https://github.com/sbaum90/AIMforGW}{\faGithub}~\cite{Baum:2023rwc}, but running in the frequency domain \texttt{AIMforGWinFspace}. 
The latter code, together with our GW signal analysis, can be found in the repository \texttt{White\_Dwarf\_Binaries\_GW} \href{https://github.com/SalaPh/White_Dwarf_Binaries_GW.git}{\faGithub} \cite{SalaWhiteDwarfBinaries2025}.
We use \texttt{PyCBC}~\cite{nitz_gwastropycbc_2024} to generate \texttt{TaylorF2} polarization basis inspiral waveforms in the frequency domain, $\widetilde{h}_{+,\times}(f)$. 
We present here a summary of this analysis, while we suggest referring to~\cite{Baum:2023rwc} for more details on the parameter estimation or~\cite{nitz_gwastropycbc_2024} for the GW signal simulator \texttt{PyCBC}.

The GW signal will be present in the frequency band the detector is sensitive to for a time much longer than it takes the detector to reorient its baseline. Then, we need to use time-dependent antenna functions as discussed in~\cite{Baum:2023rwc}. We use the leading-order time-frequency evolution $t(f)$ obtained from integrating and inverting equation~\eqref{eq:chirp2} to compute the $t(f)$-dependent antenna functions. In order to account for the changing Doppler shift of the detector relative to the source over the observation time we include a ``Doppler-phase'' $\Phi_D(f) \equiv 2\pi (\hat{\bm n} \cdot {\bm r}_D)$ where ${\bm r}_D = {\bm r}_D(t(f))$ is the time-dependent position of the detector and $\hat{\bm n}$ a unit vector in the direction of the GW source in a reference frame. We approximate the frequency-domain detector response as follows,
\begin{equation}\label{eq:FTstrain}
    \widetilde{h}(f) = e^{i \phi_D} \times \left\{ F_+[t(f)] \times \widetilde{h}_+(f) + F_\times[t(f)] \times \widetilde{h}_\times(f) \right\} ~.
\end{equation}
As discussed in~\cite{Baum:2023rwc}, this assumption neglects some of the effects of moving and rotating detectors.
We have checked explicitly that these approximations do not affect our parameter estimation forecasts for the WDB signals we are interested in here. 
We compared the new results with the ones obtained by starting from a full calculation of the detector response in the time-domain using \texttt{AIMforGW}\footnote{Equation~\eqref{eq:FTstrain} is obtained by applying the Stationary Phase Approximation (SPA)~\cite{Finn1993, Cutler1994, Poisson1995}. 
This method evaluates a fast oscillating function at the critical points of its phase and sums them up. 
While a static detector has only one critical point, a detector orbiting both the Sun and Earth can have multiple critical points due to the Doppler effect. 
The rotation around the Sun is slow and does not influence the result for the WDB of interest, but the one around Earth might. 
For this reason, the latter orbit is removed from the computation of the strain. 
We expect this suppression to be possible due to the well-known satellite trajectory.
}.

Additionally, PN corrections can be easily turned on in the code \texttt{AIMforGWinFspace}~\cite{Baum:2023rwc} up to 3.5 order. The main advantage of introducing PN corrections is that they break some degeneracies\footnote{PN corrections also add precision to the GW source evolution and strain, and consequent computations, but this improvement is negligible for most of these parameters, such as SNR.}. For example, the mass ratio $q=M_1/M_2$ influences the GW behaviour only at higher orders. Also, equation~\eqref{eq:hplus} and~\eqref{eq:hcross} show a degeneracy between $d_L$ and $\iota$, as their only entry in the strain equation is a product of each other. 
PN corrections break this $\iota-d_L$ degeneracy.
In our computation, the waveform is modelled up to PN order 3.5 for the frequency evolution and 3.0 for the amplitude.

We estimate the SNR of the signal,
\begin{equation}\label{eq:SNR}
    \rho = \sqrt{\langle h, h \rangle}~,
\end{equation} 
 via the usual inner product,
\begin{equation}\label{eq:scalarproduct}
    \langle r, s \rangle \equiv 4~{\rm Re} \int_{f_{\rm min}}^{f_{\rm max}} df\,\frac{\widetilde{r}(f)\widetilde{s}^*(f)}{S_h(f)}~,
\end{equation}
where $S_h(f)$ is the one-sided power spectral density characterising the detector's noise. 
Note that we introduced an upper cutoff $f_{\rm max}$ in the integral. 
We use this cutoff to account for the fact that the GW signal from a WDB will differ from that of two point masses assumed in the \texttt{TaylorF2} waveform model. 
Unless noted otherwise, we will use the frequency at which the lighter WD fills its Roche lobe.
Similarly, we have included a lower cutoff $f_{\rm min}$ to account for a realistic time that we can measure the signal in a detector.

The Fisher information matrix for a signal $h(t, {\bm \lambda})$ from WDB with parameters ${\bm \lambda} = \{\mathcal{M}_c, q, d_L, \ldots \}$ (see \autoref{tab:parameters}) is then obtained by computing the scalar product between derivatives of the signal with respect to the parameters,
\begin{equation}
    \Gamma_{ij} \equiv \left\langle \frac{\partial h}{\partial \lambda_i}, \frac{\partial h}{\partial \lambda_j} \right\rangle~,
\end{equation}
where again, we included an upper cut $f_{\rm max}$ in the computation of the inner product. The variance-covariance matrix is obtained by inverting the Fisher information matrix, $\mathcal{C}_{ij} \equiv (\Gamma^{-1})_{ij}$.
Note that, as in~\cite{Baum:2023rwc}, we apply priors on each angle by adding $1/\pi^2$ ($1/2\pi^2$) to the diagonal entries of the Fisher matrix corresponding to angles with range $\pi$ ($2\pi$) in order to restrict these parameters to their physical ranges. 
The uncertainty with which a parameter can be measured is then approximated by 
$\sigma_i \equiv \sqrt{\mathcal{C}_{ii}}$.

\subsection{Parameter estimation}\label{sec:GWsignal}

MAGIS Space is expected to detect a multitude of WDBs, far away and close to the end of their lifetime. Following a Fisher analysis, as described above, we can simulate the GW signal from these binaries with \texttt{AIMforGWinFspace}~\cite{Baum:2023rwc} and estimate the precision with which the detector will measure their different parameters (see \autoref{tab:parameters}) that characterize their waveforms.

In order to show these results, we fix the time frame of observation as follows.
We set a frequency cutoff for the end of the measurement in equation~\eqref{eq:scalarproduct} at the beginning of RLOF, equation~\eqref{eq:RLOFfreq},
\begin{equation}
    f_{\rm max} = f_{GW,RLOF}~.
\end{equation}
Up to this moment, WDB can still be considered point masses, the binary evolution is still described by GW radiation only, and its waveform is simplified as effects after RLOF do not need to be considered yet (see subsections \ref{sec:RLOF} and \ref{sec:MT}).
However, as a binary approaches merger and its gravitational wave frequency increases, the signal becomes stronger, allowing for more precise parameter estimation.
Indeed, as discussed in \autoref{sec:MT}, the GW signal from a WDB will be practically identical to that of two point masses for a considerable amount of time after the onset of Roche lobe overflow.
Hence, this cutoff is a rather conservative approximation.
We carry out this estimation for a realistic observation time of the detector of one year, meaning that $f_{\rm min}$ in equation~\eqref{eq:scalarproduct} is evaluated $1$ year before $f_{\rm max}$.

\begin{table}[t]
    \centering
    \begin{tabular}{c|l|c}
    Symbol & Parameter & Value\\
    \hline
        $f_{\rm max}$ & Cutoff frequency & $f_{GW,RLOF}$\\
        $\Delta t_{obs}$ & Observation time & $1$\,yr\\
        $d_L$ & Luminosity distance & $25.0$\,Mpc\\
        $\Phi_0$ & Binary phase at $t_0$ & $0^\circ$\\
        $t_0$ & Time of reference & $0$\,s\\
        $\iota$ & Inclination angle & $45^\circ$\\
        $\psi$ & Polarization angle & $135^\circ$\\
        $\alpha_0$ & Right ascension & $180^\circ$\\
        $\delta_0$ & Declination & $-30^\circ$\\
    \end{tabular}
    \caption{Parameters used for the Fisher analysis using \texttt{AIMforGWinFspace}. }
    \label{tab:fixedparam}
\end{table}

Regarding the parameters describing the GW signal, we perform the computation for each combination of the two masses $M_1$ and $M_2$ (or equivalently $\mathcal{M}_c$ and $q$), while the other parameters are fixed\footnote{The parameters $\Phi_0$ and $t_0$ are set to $0$. Modifying them only changes the initial phase of the binary but does not influence the results.}. We select a case study which is representative of the various detectable WDB. 
In particular, $d_L=25.0$\,Mpc since it represents the region in which we detect the most WDB that might go SN Ia with MAGIS Space, as it will be shown in \autoref{sec:mergerrate}.
A summary of the fixed parameters for the following analysis can be found in \autoref{tab:fixedparam}.

\begin{figure}[t]
    \centering
    \includegraphics[width=0.85\linewidth]{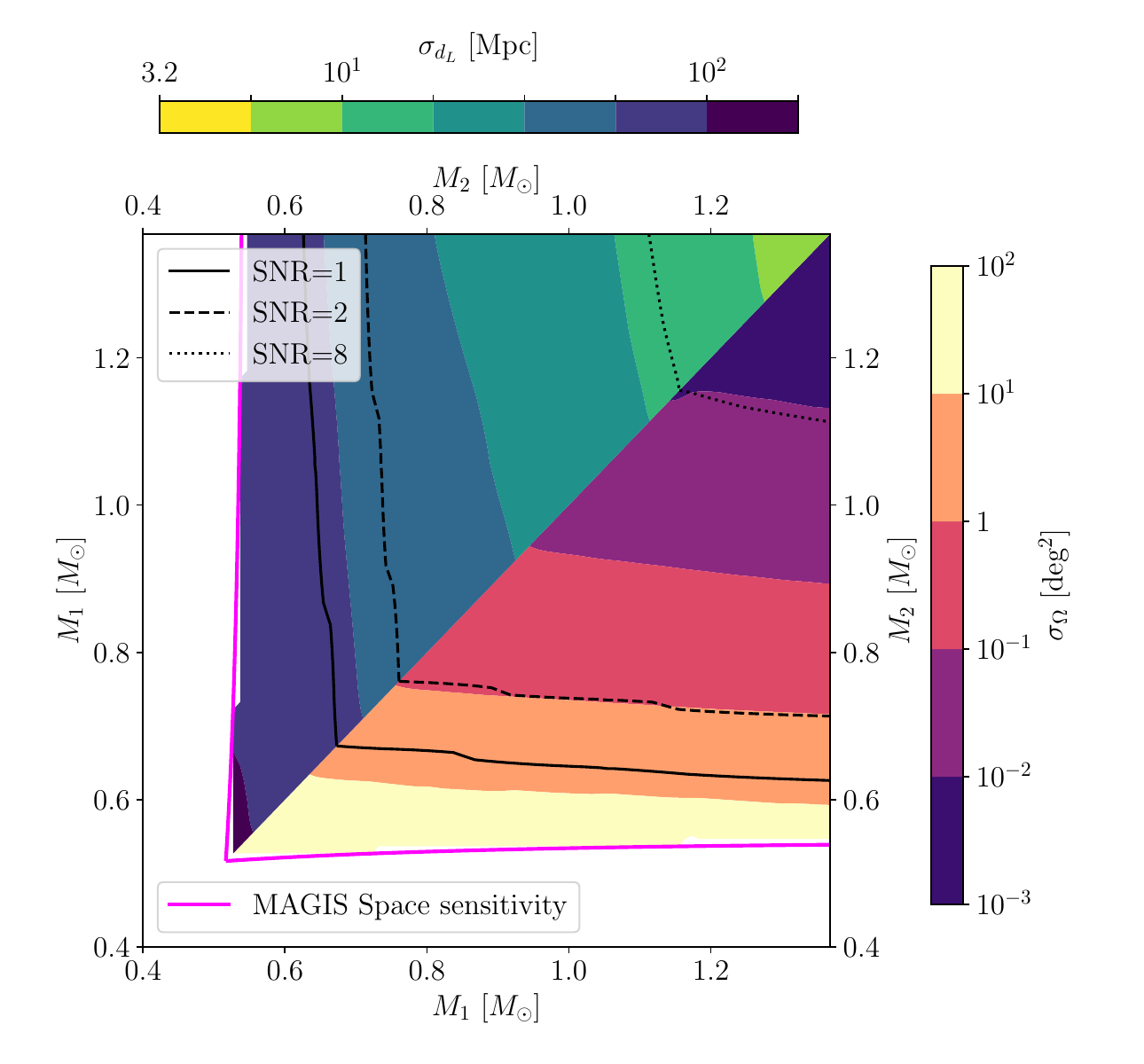}

    \caption{Results of the Fisher analysis for the uncertainty on sky localisation $\Omega$ at the bottom and $d_L$ in the upper part, computed for fixed parameters summarised in \autoref{tab:fixedparam}. The sky position can be found from parameters $\alpha$ and $\delta$ with equation (\ref{eq:skylocalisation}). The computation considers a GW evolution only of the binaries for $1$\,yr before RLOF. The black lines show the lowest mass combination that reaches an SNR of $1$, $2$ and $8$ for the same observation. We apply Gaussian smoothing to erase numerical noise.}
    \label{fig:MeasurementPrecision_dL_Sky}
\end{figure}

\begin{figure}[t]
    \centering
    \includegraphics[width=0.85\linewidth]{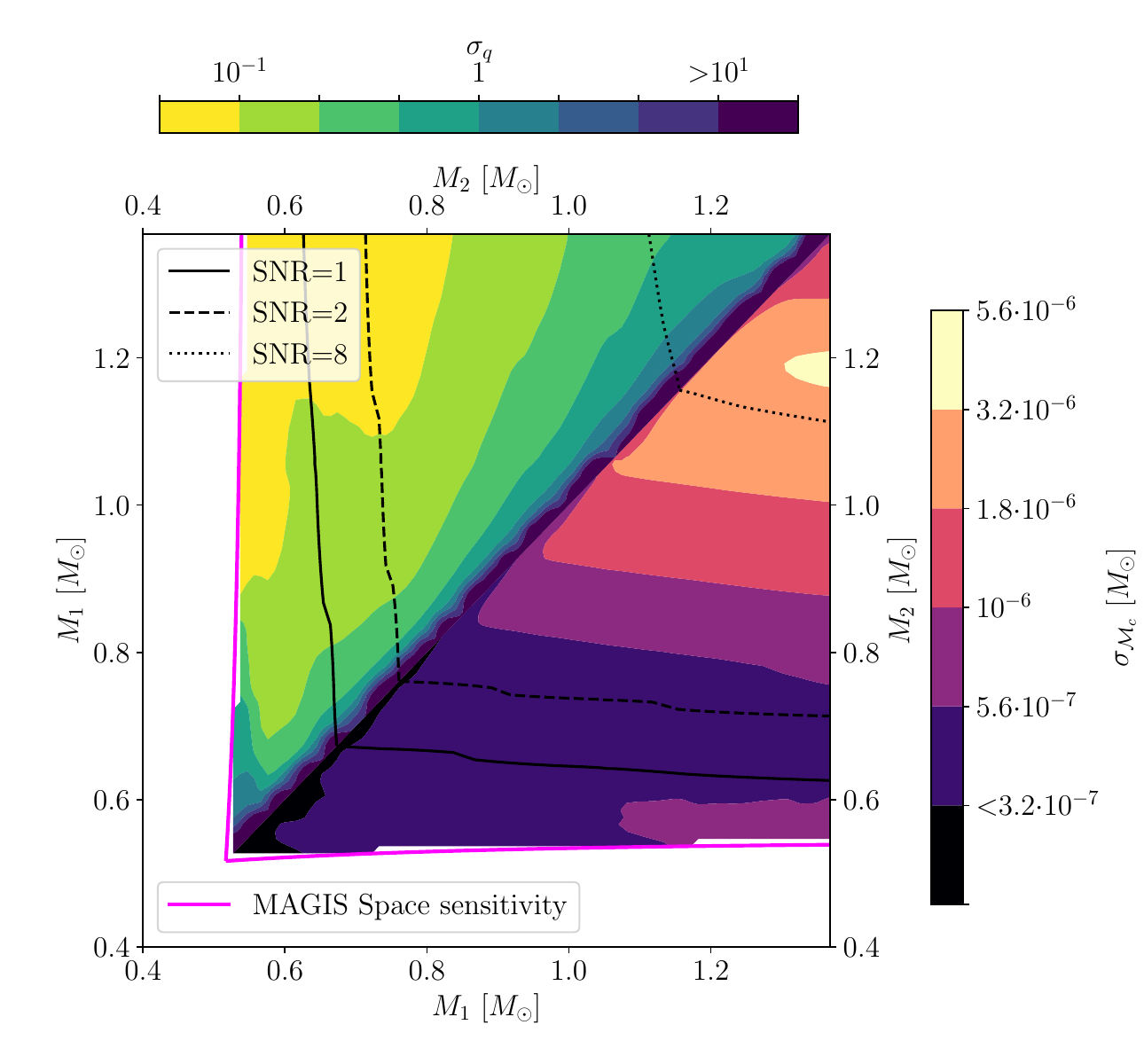}

    \caption{Results of the Fisher analysis for the uncertainty on $\mathcal{M}_c$ at the bottom and $q$ in the upper part, computed for fixed parameters summarised in \autoref{tab:fixedparam}. The computation considers a GW evolution only of the binaries for $1$\,yr before RLOF. The black lines show the lowest mass combination that reaches an SNR of $1$, $2$ and $8$ for the same observation. We apply Gaussian smoothing to erase numerical noise.}
    \label{fig:MeasurementPrecision_q_Mc}
\end{figure}

In Figures \ref{fig:MeasurementPrecision_dL_Sky} and \ref{fig:MeasurementPrecision_q_Mc}, we show the expected precision on $4$ important parameters that MAGIS Space can achieve.  
\autoref{fig:MeasurementPrecision_dL_Sky} reports the error on the luminosity distance ($d_L$) and the sky localisation in terms of solid angle (${\Omega}$)~\cite{PhysRevD.57.7089}; from the entries of the variance-covariance matrix ($\mathcal{C}_{ij}$) related to the binary's right ascension ($\alpha$) and declination ($\delta$), the uncertainty on $\Omega$ is computed via
\begin{equation}\label{eq:skylocalisation}
    \sigma_\Omega = 2 \pi \cos\delta \cdot \sqrt{\mathcal{C}_{\alpha\alpha} \mathcal{C}_{\delta\delta} - \mathcal{C}_{\alpha\delta}^2} \;.
\end{equation}
\autoref{fig:MeasurementPrecision_q_Mc} shows the estimated precision on the mass relation ($q$) and the chirp mass ($\mathcal{M}_c$). 
Both plots include a threshold showing the lowest mass combinations that reach an SNR of $1$, $2$ and $8$ in 1 year of observation before RLOF. 
We consider a WDB to be detectable only if the SNR is higher than $2$.

A few considerations can be made based on these results. 
Firstly, most WDBs will not be very loud at $d_L=25.0$\,Mpc. Only binaries with at least one relatively massive component are going to be detectable. 
While this low sensitivity to lighter WDs limits the number of total detections, it represents a less serious problem in terms of binaries that go SN Ia, as they need larger masses. 
In addition, the SNR \eqref{eq:SNR} scales inversely with the luminosity distance, $d_L$.
From \autoref{fig:MeasurementPrecision_dL_Sky}, we can notice that MAGIS Space is not expected to be very precise in measuring the luminosity distance. 
Only the distance of the most massive binaries can be measured with significant precision, below $10$\,Mpc. 
Concerning the sky localisation, our predictions show high accuracy, with an uncertainty on the solid angle of $1$\,${\deg}^2$ for the lightest detectable binaries (with SNR higher than $2$), up to better than $10^{-2}$\,${\rm deg}^2$ for the most massive ones. 
This is a very promising result for the observation of SN Ia events from the DD scenario following a GW detection and merger forecast.

The accuracy of the measurement of the masses of the binaries is described in \autoref{fig:MeasurementPrecision_q_Mc} with the GW parameters $\mathcal{M}_c$ and $q$. These parameters are a more suitable choice for reconstruction than the component masses $M_1$ and $M_2$ as their impact on the waveform is fundamentally different, see equations~\eqref{eq:hplus} and~\eqref{eq:hcross}. While the chirp mass plays a primary role in modelling the GW signal, $q$ only enters the strain equation at higher PN orders. 
These different contributions could anticipate the results of \autoref{fig:MeasurementPrecision_q_Mc}, which shows very high precision on the measurement of $\mathcal{M}_c$, always better than $10^{-5}\,M_\odot$, and a poor one for the mass ratio. 
Because of their formation history, most binaries have a value of $q$ between $1$ and $5$, rarely reaching larger values. 
Consequently, the estimated uncertainties appear quite large, much higher than $10\,\%$ for most detectable mass combinations, with SNR higher than 2.

An important outcome of this analysis concerns the possibility of predicting the WDB merger based on the reconstructed parameters of the GW source. 
Since we stop our observation at RLOF and not at coalescence, we compute the prediction of the instant a binary reaches RLOF with the derived variables, obtaining very poor outcomes. 
The moment a binary reaches RLOF cannot be predicted with a better uncertainty than $1$\,yr, for any mass pairing. 
This inaccuracy comes as a consequence of the strong dependencies on various parameters in the computation of Roche lobe radius, such as on $\mathcal{M}_c$, $t_0$ and especially $q$, see equation~\eqref{eq:RocheLobe}. 
The poor prediction of RLOF is followed by an even more inaccurate forecast of the merger due to the uncertainty in the theoretical modelling of mass transfer. In \autoref{sec:Mergersignal} we thus discuss an alternative method for detecting an upcoming WDB merger from the GW signal alone. 
\\[1\baselineskip]
In this section, we explained the method used to predict the performance of MAGIS Space and found that one could be optimistic in localising WDBs in the sky to eventually observe a SN Ia event. For a more detailed discussion of how the SNR and parameter dependency change with the parameters of the binary, we suggest reading~\cite{Baum:2023rwc}.
In \autoref{sec:mergerrate}, we consider the dependencies on all parameters to compute the detectable population of WDBs.

\section{Merger detection}\label{sec:Mergersignal}

Several reasons make predicting WDB mergers and their subsequent eventual detonation difficult, including: the moment a binary reaches RLOF cannot be reconstructed with precision, the dynamics after mass transfer starts are not precisely known, the final event is not unique and there are large uncertainties on many parameters as the WDB's GW signal is not expected to be very loud. 
Simulations have been performed to study specific cases~\cite{Loren-Aguilar:2009iym,Pakmor_2010,Pakmor_2012,Raskin_2012,Ji_2013,Benz:1990,Dan:2011aw,Dan:2012dt,Raskin:2013wqa}. 

\begin{figure}[t]
\centering
\includegraphics[width=0.85\linewidth]{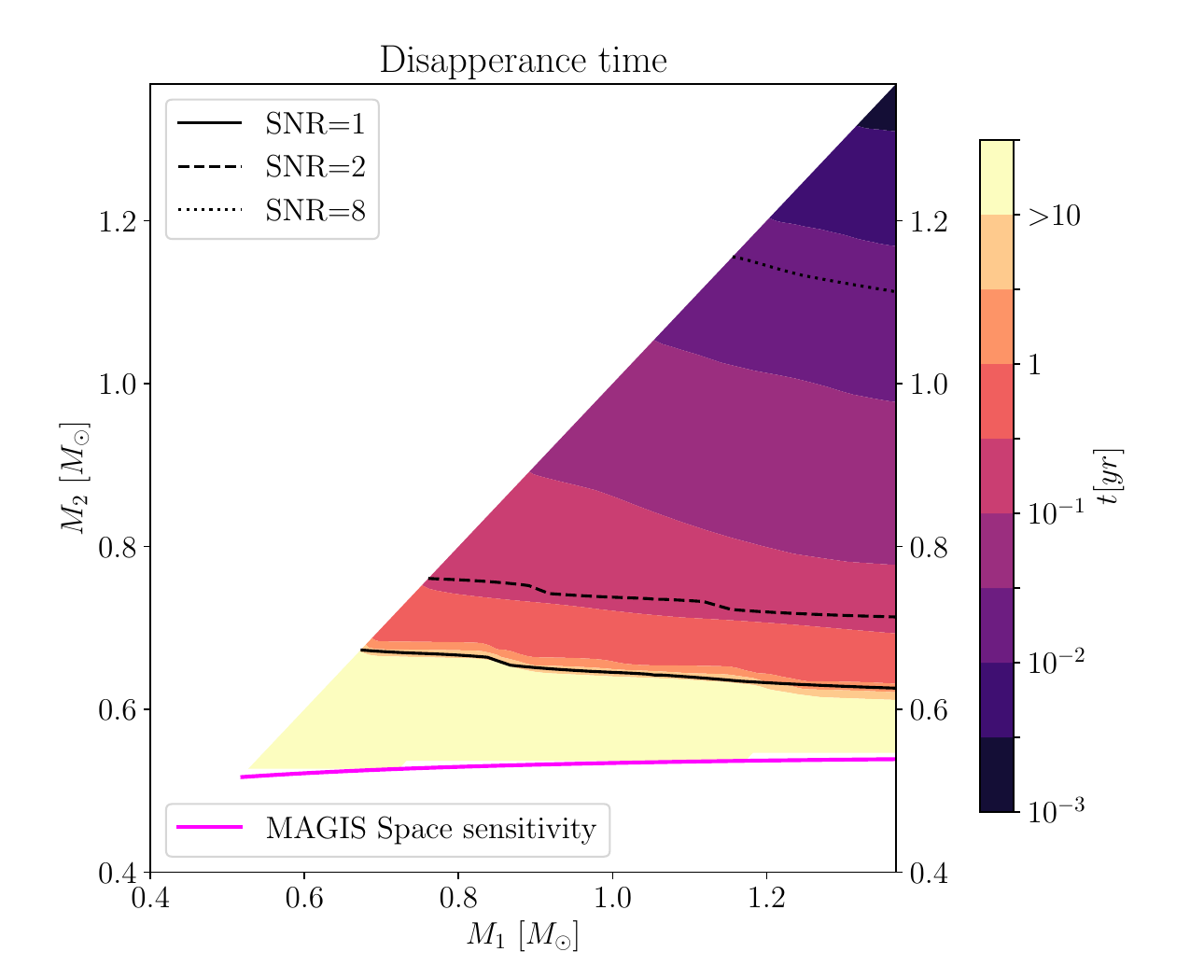}
\caption{Time after RLOF required to realise the WDB has merged for binaries with parameters described in \autoref{tab:fixedparam}. 
The black lines show the lowest mass WDBs that could already be observed with an SNR of $1$, $2$ and $8$ for one year of observation before RLOF, as already displayed in Figures \ref{fig:MeasurementPrecision_dL_Sky} and \ref{fig:MeasurementPrecision_q_Mc}. We apply Gaussian smoothing to erase numerical noise.}
\label{fig:SignalDisapperance}
\end{figure}

We develop an effective concept to identify and observe the final events of WDBs, which is based on GW measurement and the duration of SN Ia afterglow.
We start by assuming that we can detect a WDB and reconstruct its parameters, analogously to \autoref{sec:GWsignal}, but have no information regarding an eventual imminent coalescence.
First, we calculate how quickly MAGIS Space will be able to recognise that a binary has merged, meaning that there is no more GW signature.
If this signal is detected within a time span shorter than the SN Ia remnant’s luminous lifetime, on the order of a few weeks, then an EM counterpart from the merger may still be observable.
Second, we can identify the onset of mass transfer and predict the coalescence by analysing the phase difference between the theoretical model exclusively driven by GW and the observed signal.
We take a phase difference of $\pi$ to represent an alert for mass transfer, meaning that WDB is approaching a merger.
The chirp mass $\mathcal{M}_c$ solely determines the frequency evolution of an ideal point-mass binary, see equation~\eqref{eq:chirp2}, and in general, predictions based on this mass parameter should be reliable, as it can be reconstructed with high accuracy (see \autoref{fig:MeasurementPrecision_q_Mc}).

The first question is whether it is possible to recognise that a binary has merged through the disappearance of the GW signal.  We find the time necessary to realize this by computing the period after RLOF required to measure an SNR of $1$ with a purely GW-dominated evolution\footnote{Here we use a GW binary evolution after RLOF, without other effects, as this is the signal we expect to measure if the binary is not made of WDs but of black holes or neutron stars instead (point-mass approximation).
 }. 
Starting the computation at RLOF is a conservative approximation and provides a natural counterpart to the signal analysed in \autoref{sec:GWsignal}\footnote{Clearly, the onset of RLOF does not correspond to the merger itself. Considering the actual merger is complicated, but would lead to a shorter time to reach an SNR of $1$. Note that the longer a binary lives, the closer the two compact objects get, the higher frequencies it reaches, and the louder the signal becomes.}.
\autoref{fig:SignalDisapperance} shows the results for the representative case discussed in above, with the parameters summarised in \autoref{tab:fixedparam}. 
If this signal is not detected, the binary must have stopped emitting GWs. 
The disappearance of WDBs' signal with $\mathcal{M}_c\gtrsim0.9\,M_\odot$ is detectable within a month or less at a distance of $25.0$\,Mpc, where the connection with the chirp mass can be established from \autoref{fig:WDB_RLOF}.
These timeframes are very promising for observing the aftermath of WDBs' final events.
On the other hand, it is clear from \autoref{fig:SignalDisapperance} that if a binary remains unidentified prior to RLOF, its eventual disappearance will also go undetected.
As previously mentioned, this reasoning needs to be rescaled at different distances.

\begin{figure}[t]
\centering
\includegraphics[width=\linewidth]{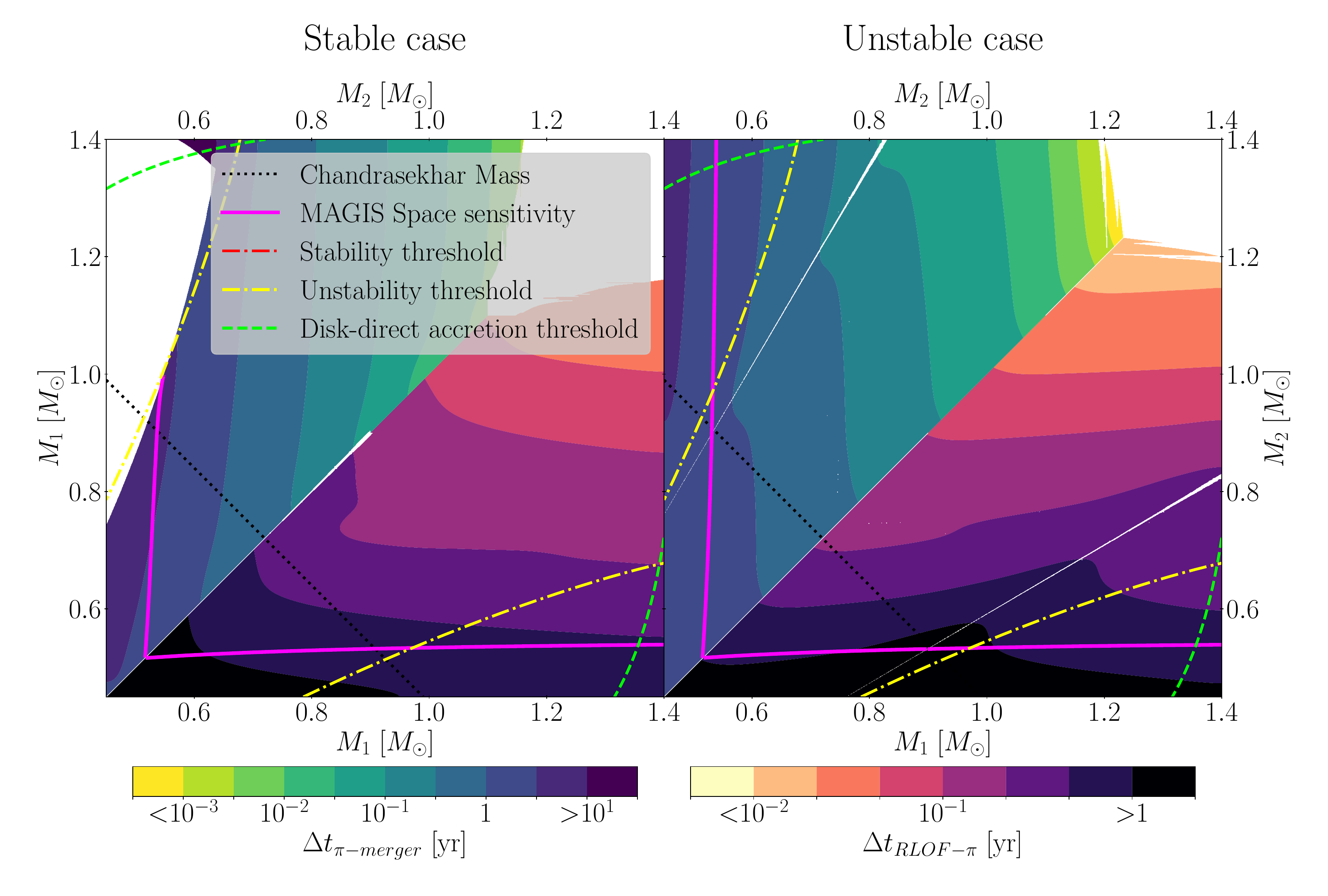}

\caption{On the bottom part of the plots, the time interval from RLOF to $1\pi$ of phase difference between point mass binary and WDB with mass transfer is shown. On the upper part,  the time interval for mass-transferring WDBs between the latter moment and the merger (when mass transfer becomes unstable) is displayed. We show the two extreme cases, the most stable scenario on the left and the most unstable on the right, see \autoref{sec:MT}. In addition, the same unstability threshold and accretion disk limit are displayed here, as in \autoref{fig:WDB_RLOF}. It is highlighted how the stability threshold lies outside of the mass range considered. We apply Gaussian smoothing to erase numerical noise. }
\label{fig:RLOF1pi}
\end{figure}

While the concept of recognising a merger in its aftermath is straightforward, its prediction requires a more involved approach. This forecast can be achieved by identifying a mass-transferring process in WDB that would alert us of an imminent merger. 
If a WDB has been observed for long enough before it reaches RLOF, its waveform $h_{GW}$ (determined by GW emission only as in \autoref{sec:parameters}) can be reconstructed. 
The mass transfer progressively perturbs the binary's evolution, as described in \autoref{sec:MT}, leading to the emission and detection of a modified waveform $h_{\rm obs}$. 
In particular, the two oscillations will start to differ. While the reconstructed and purely GW-dominated waveform fluctuates as $\Phi_{GW}(t)$, defined in equation~\eqref{eq:phase}, the waveform with mass transfer will accumulate a phase $\Delta\Phi_{\rm obs}(t)$ over time,
\begin{equation}
    \Phi_{\rm obs}(t) = \Phi_{GW}(t) + \Delta\Phi_{\rm obs}(t)~.
\end{equation}
The two strains can be constantly monitored during observation. The phase dependence of the overlap integral, computed using the scalar product in equation~\eqref{eq:scalarproduct}, is the following,
\begin{equation}
   O(h_{GW}, h_{\rm obs}) = \frac{\langle h_{GW} , h_{\rm obs} \rangle}{\sqrt{\langle h_{GW} , h_{GW} \rangle \langle h_{\rm obs} , h_{\rm obs} \rangle}} \approx H(t) e^{i\Delta\Phi_{\rm obs}(t)}~.
\end{equation}
The coefficient $H(t)$ is determined by the two $h_{GW}$ and $h_{\rm obs}$ strain amplitudes and varies slowly. The overlap integral time dependence is predominantly governed by the phase difference $\Delta\Phi_{\rm obs}(t)$.
For more compact objects' binaries, that might be black hole binaries or neutron star binaries, the evolution is unperturbed: $O(h_{GW}, h_{\rm obs})=1$. 
In this particular case, no alert of an imminent merger is detected, and the binary continues to evolve instead. These systems are just as interesting WDBs, since they may later evolve into sources observable in higher frequencies in the gravitational spectrum.
As soon as the binary evolution deviates from the point-mass description, the overlap integral decreases with growing phase difference.
The first strong evidence of mass transfer is represented by the overlap integral reaching its minimum, namely $\Delta\Phi_{\rm obs}(t)=\pi$.

We model the evolution of mass-transferring binaries, as outlined in \autoref{sec:MT}, to determine the timescales for the earliest potential merger warning signal and the merger itself, considering both the most stable and unstable cases.
More specifically, \autoref{fig:RLOF1pi} reports the period between RLOF (\ref{eq:RocheLobe}) and the moment a mass-transferring WDB cumulates a phase difference of $\pi$, $\Delta t_{RLOF-\pi}$, and also the interval between this latter point in time and the merger, $\Delta t_{\pi-merger}$. 
The sum of the two times gives the period between RLOF and the merger. 
The two limiting scenarios represent the longest and shortest possible lifetime of each binary\footnote{We did not account for the disk-direct accretion effect. An impact of this approximation is visible along the top-left edge of the plot.}.
A realistic outcome is expected to fall between the two. 
We notice that almost all WDBs, detectable with MAGIS Space, cumulate a phase difference of $\pi$ within a year after RLOF\footnote{The diagonal split in $\Delta t_{RLOF-\pi}$ plot for the unstable case represents the threshold between positive and negative phase difference. Above this split, the $\Delta\Phi_{\rm obs}(t)$ is positive, below is negative. Such an exact alignment of masses is extremely rare, but would lead to a mass-transfer that does not dephase, hence cannot be recognised.
For the stable case $\Delta\Phi_{\rm obs}(t)$ is always negative.}.  
The ratio between $\Delta t_{RLOF-\pi}$ and $\Delta t_{\pi-merger}$ decreases from about $10$ for the lightest observable WDBs, where timescales are on the order of a year, to approximately $1$ at $\mathcal{M}_c\sim1.0\,M_\odot$, where the timescales shrink to at most a few weeks.
Binaries with $\mathcal{M}_c\gtrsim1.1\,M_\odot$ reach higher frequencies at RLOF due to their compactness; thereafter, the stronger GW emission drives the evolution to a quick merger, before reaching a phase difference of $\pi$. 
The merger of the most massive WDBs is thus hard to predict, as they cannot be distinguished from more compact objects\footnote{For example, the range of masses $1.1-1.4\,M_\odot$ might easily represent binary neutron stars. Before merger, the GW signals from WDBs and neutron star binaries will be extremely similar, preventing us from differentiating the two systems. In principle, EM telescopes could observe the binary prior to merger and differentiate between WDs and neutron stars, however, such observations are very challenging due to the faintness of WDs and most neutron stars.}.  

Mass transfer effects on strain evolution could potentially have been degenerate with other parameters, which would make it difficult to identify binaries already exchanging matter at detection.
We modelled WDB evolution for several test cases, including mass transfer (\autoref{sec:MT}) and verified that it does not exhibit degeneracy with other waveform parameters.
As a consequence, although quantifying proximity to merger is more challenging when observing a WDB already undergoing mass transfer, we can still accurately identify the system and reconstruct its GW parameters.
\\[1\baselineskip]
Being able to observe the EM signal from the merger of WDB is a challenging task. In the previous \autoref{sec:GW}, we discussed how well MAGIS Space can localize WDBs. 
In this section, we have shown two possibilities to either forecast the coalescence of WDB or quickly recognise it, to then observe the subsequent SN Ia, the lack of it or other outcomes.
We developed a strategy to achieve this goal for any WD mass combinations.
For WDBs with $\mathcal{M}_c\lesssim 1.0\,M_\odot$ merging within the MAGIS Space frequency band, the waveform dephasing due to mass transfer can provide an early warning for coalescence from a few days to a year in advance (see \autoref{fig:RLOF1pi}).
For the more massive binaries, matter exchange cannot be identified, but the signal disappearance after the merger can be detected promptly, and the aftermath of coalescence can still be observed. 
At $d_L=25.0$\,Mpc, the lack of GW emission for WDBs with $\mathcal{M}_c\gtrsim 0.95\,M_\odot$ can be noticed approximately within $15$ days or less (see \autoref{fig:SignalDisapperance}).
Overall, a follow-up EM observation is possible only if the binaries and their dephasing are detectable by MAGIS Space, hence reach an SNR of at least $1$ both before and after RLOF\footnote{This means that the GW signature of a WDB must reach an SNR of $1$ within the time a mass-transferring binary takes to reach a phase of $\pi$.}.
Due to the SNR-$d_L$ linear relation, this condition depends not only on the WD masses but more importantly on the source luminosity distance. 
Ultimately, we can confidently expect to observe multi-messenger signals from all WDB mergers detectable by MAGIS Space.

\section{Merger rate}\label{sec:mergerrate}

We now turn our attention to quantifying the number of future detections of WDB mergers and SN Ia that are possible with MAGIS Space.
We begin by clarifying the assumptions underlying the terms merger and detection in this section. Analogously to the analysis in sections \ref{sec:GW} and \ref{sec:Mergersignal}, we base our analysis on the GW signal with point mass approximation, hence before RLOF. We assume that every binary that reaches the RLOF is destined to merge.
It is worth noting that a small portion of WDBs, if configured in a particularly stable way, may have significantly extended lifetimes, see discussion in \autoref{sec:MT}. 
However, these binaries mostly lie outside of the sensitivity of MAGIS Space, see \autoref{fig:RLOF1pi}, and when they do not, they have an uncommonly large $q$. 
Their tiny contribution can be checked in \autoref{app:contributions}.
The uncertainties regarding the stability of WDBs, combined with the small fraction of the population affected by this correction, justify our assumption. 
Additionally, from the previous sections \ref{sec:GW} and \ref{sec:Mergersignal} we determined that every WDB merger can be identified, either predicted or quickly recognised a posteriori, as long as its signal is loud enough.
Hence, we consider that an atom-interferometer can detect a WDB and its merger if the GW signal reaches an SNR of $2$ within one year of observation prior to RLOF.

The last ingredient missing to evaluate the detectable mergers and SN Ia rates is an estimate of the WDB population.
To be conservative and account for a range of possibilities, we adopt two different methods.
\begin{enumerate}[label=\roman*.]
    \item \underline{Model approach}: a theoretical model based on the star formation history and described in \autoref{sec:WD}, is used to compute the population synthesis of WDB~\cite{Seppe24,Hofman24}. Consequently, both the mass function of WDB and the total WDB number density can be estimated.
    
    \item \underline{SN Ia rate approach}: while the WDB number density is not known from observation, the SN Ia rate is. 
    Assuming some mass combinations that lead to SN Ia (see \autoref{sec:SNformation}), knowing a WDB mass function from the model approach and making some assumption on the contribution from the DD scenario, it is then possible to go from the SN Ia rate to a WDB population. 
    Hence, we can relate the WDB merger rate to observation.
    In the results, we use the notation $DD\%$ to indicate the percentage of SN Ia originating from the DD scenario. 

\end{enumerate}

This section is organised as follows: in \autoref{sec:WDBpop} we reconstruct the WDB population distribution in terms of mass combinations from the star formation history, which is required for both approaches. In \autoref{sec:SNIarate} the observed SN Ia rate, required for the SN Ia method, is introduced. 
In \autoref{sec:WDBdetections}, we describe how to filter only the detectable binaries over different distances and parameters by using 
the tool described in \autoref{sec:GWsignal}.
The last \autoref{sec:results} presents the results on the number of WDB mergers we expect to see with MAGIS Space, and also the possibility of detecting SN Ia. Secondarily, AEDGE capabilities are also discussed.

\subsection{White Dwarf Binary mergers population}\label{sec:WDBpop}

Of all WDB merger events, only a fraction can be detected, and an even smaller portion can be both detected and go SN Ia from the DD scenario.
Calculating these fractions represents the first step required to find the overall WDB merger rate, which can be achieved with the assumptions made in \autoref{sec:SNformation} about the binaries that produce SN Ia, see~\eqref{eq:SNconservative} and \eqref{eq:SNoptimistic}, 
and the WDB population synthesis.
Attempts to compute the latter component have been made~\cite{Nelemans_2001_galax,Marsh_2011,Badenes:2012ak,Toonen_2012,Rebassa_Mansergas_2018,  Maoz:2018epf,Breivik_2020,DWDmasses}. 
Nevertheless, this information cannot be computed from observation due to the lack of data, hence must be inferred from theoretical assumptions on WDB formation, GW evolution and final phases, described in \autoref{sec:WDBevolution}. 
Articles~\cite{Seppe24,Hofman24} tackle this challenge by simulating the formation and evolution of WDBs. 
The result is a set of different populations of WDB, with respective times to RLOF, which vary depending on model, metallicity and star formation rate densities and evolution, as mentioned in \autoref{sec:WD}.
We use the populations provided by these studies to evaluate the outcome we need. 
More specifically, the outcome of $12$ combinations of different star formation rate histories and binary formation models is used to estimate the uncertainties.
We define a fiducial WDB population model, which is based on the star formation densities from~\cite{Chruslinska2019} with a $\gamma\alpha$ binary model with $\gamma=1.75$ and $\alpha=4$~\cite{Hofman24}.

\begin{figure}[t]
\centering
\includegraphics[height=0.7\linewidth]{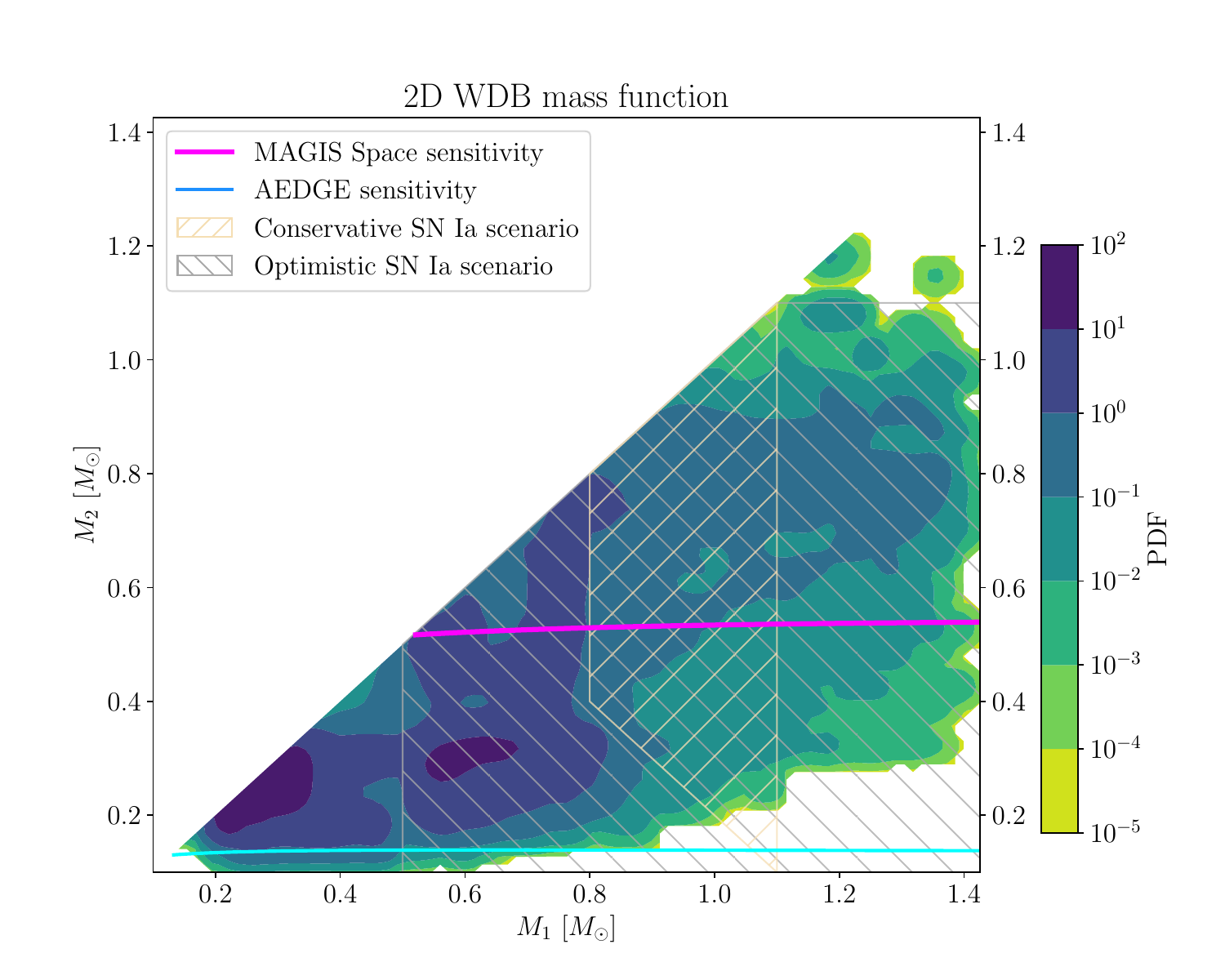}

\caption{2D probability density function (PDF) of WDBs merging today (in units of $M_\odot^{-2}$). The PDF is recovered from a simulated population based on the star formation model $\gamma\alpha$ ($\gamma=1.75$ and $\alpha=4$) and weighted over different metallicities and redshifts. MAGIS Space and AEDGE sensitivity lines indicate the lowest mass combinations that reach the lowest measurable frequency of the two detectors at RLOF. Note that, only $13\,\%$ of all WDBs are potentially detectable by MAGIS Space (above its sensitivity line), while this portion goes up to $99.8\,\%$ for AEDGE.
We apply Gaussian smoothing to erase numerical noise. }
\label{fig:WDBweighted}
\end{figure}

The population of WDBs merging today is based on the sum of the binaries forming with different metallicities, weighted by their respective star formation rates and evaluated on their formation history, which is dependent on their distance.
The population synthesis we obtain for the fiducial model is shown in \autoref{fig:WDBweighted}, where we also overlay the two mass regions that lead to SN Ia, see equations~\eqref{eq:SNconservative} and~\eqref{eq:SNoptimistic}. 
In terms of detections with MAGIS, only the binaries above the pink line are of interest, while for AEDGE (red line), almost every merger is detectable.
It must be mentioned that the results from~\cite{Seppe24,Hofman24} give the period from formation to RLOF, while we are looking for the merging population. In the following, we neglected this difference since the expected time between RLOF and merger, as shown in \autoref{fig:RLOF1pi}, is tiny compared to the stellar formation timescales of Gyr, discussed in \autoref{sec:WDBevolution}.

The total number of WDB can be computed directly from the reference~\cite{Seppe24,Hofman24}, which provides the star formation rate density for each metallicity, thus the merging WDB spatial distribution.

\subsection{Observed SN Ia rate}\label{sec:SNIarate}
The second approach to infer the detectable WDB merger rate is based on the observation of SN Ia events. 
The fact that only some unknown fraction of SN Ia originates from the DD scenario adds a layer of uncertainty. 
Nevertheless, starting from well-measured quantities such as the volumetric SN Ia rate is expected to provide us with a good estimate of the expected WDB merger rates.

Many projects studied SN Ia detections over the years~\cite{Magdwick_2003,Panagia_2007,Krughoff_2011,Maoz_2012,Ruiter:2012ys,Graur_2013,Kistler_2013,Maoz_2014,Brown_2019,Frohmaier2019,Zwicky2020,Sharon2021,Wiseman_2021}. 
We refer to some of the more recent results from Palomar Transient Factory and Zwicky Transient Facility, given in~\cite{Frohmaier2019,Zwicky2020,Sharon2021}.
The SN Ia volumetric rate is
\begin{equation}\label{eq:SNIarate}
    r_{SNIa}=2.43^{+1.10}_{-0.32}\times 10^{-5}\ \text{yr}^{-1}\text{Mpc}^{-3}~,
\end{equation}
where the central value is taken from~\cite{Frohmaier2019} and the errors from the extreme values in~\cite{Zwicky2020,Sharon2021}. 
This value is used as the starting point to compute WDB total mergers and detectable ones depending on the population synthesis (\autoref{fig:WDBweighted}) and detector capabilities, for the SN Ia rate approach. 
This rate is then reported in \autoref{fig:Mergerrates}.

\subsection{Detectable rates}\label{sec:WDBdetections}
\begin{figure}[!tp]
    \centering
    \includegraphics[width=\linewidth]{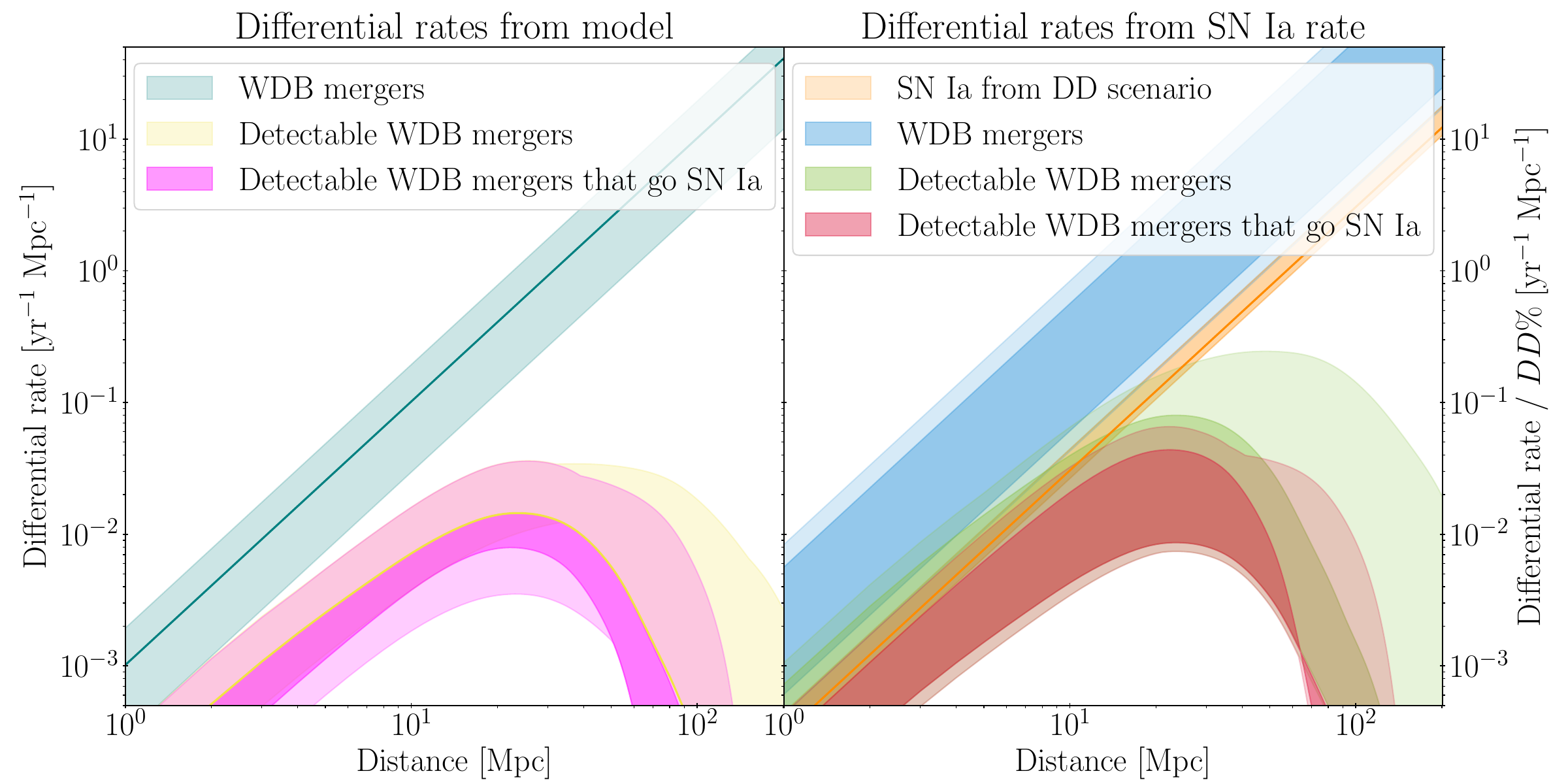}

    \includegraphics[width=\linewidth]{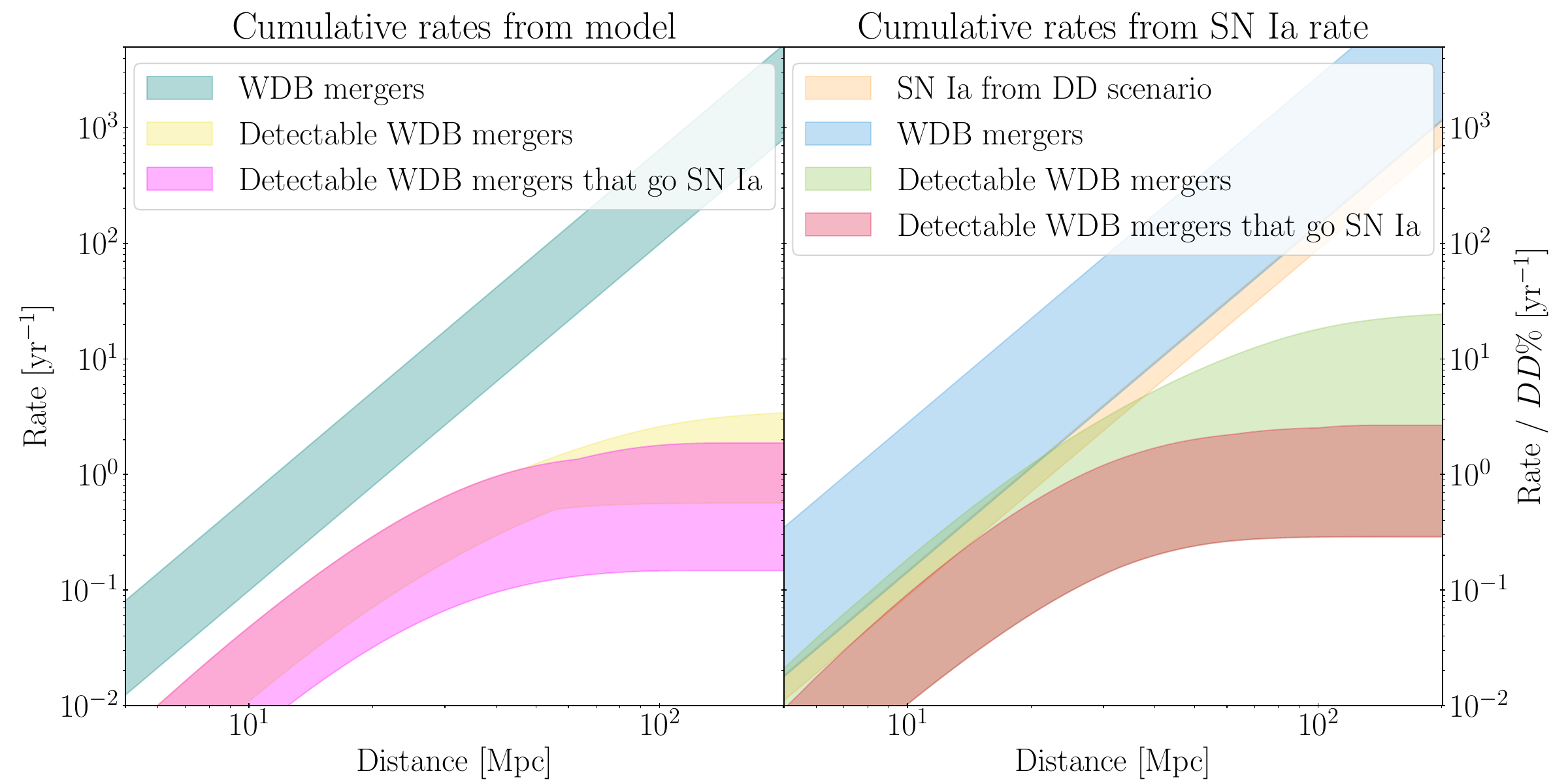}
    \caption{Total WDB merger rates and rates for mergers detectable by MAGIS Space as a function of the luminosity distance. 
    The top figures shows the differential rates versus distance, while the bottom shows the cumulative rates within that distance. The left plots show results using the theoretical model approach for WDB population synthesis, while the rates on the right plots are computed starting from SN Ia observations. 
    All the results on the right should be scaled with the fraction of SN Ia that originate from the DD scenario ($DD\%$). 
    A binary merger is defined as detectable if its SNR for one year of observation before RLOF is higher than $2$ (see dashed line in Figures \ref{fig:MeasurementPrecision_dL_Sky}, \ref{fig:MeasurementPrecision_q_Mc} and \ref{fig:SignalDisapperance} for a specific case). In all plots, the darker shape shows the fiducial values given by the interval between the two extreme SN Ia production cases (see \autoref{sec:SNformation}) in combination with the fiducial star formation model $\gamma=1.75$, $\alpha=4$ from \autoref{sec:WD}. Solid lines instead of bands are plotted when the fiducial values are singular and do not span a range. The wider and lighter bands show a larger range of possibilities, including uncertainties and different star formation models. The cumulative plots only show the latter bands.}
    \label{fig:Mergerrates}
    \vfill
\end{figure}

The two previous procedures allow us to compute the total expected WDB merger rates; directly from the theoretical population synthesis using the model approach, and by starting from SN Ia rates and applying the assumptions made on the fraction of merger that go SN Ia, from \autoref{sec:SNformation}, and the expected portion of SN Ia that come from DD scenario. 
This information is also reported in \autoref{fig:Mergerrates} for both methods (model approach on the left and the SN Ia rate approach on the right plots).

The next logical step is to estimate the rates of the detectable merger and SN Ia, hence compute which WDB coalescence can be observed. 
For this to happen, the GW signal mentioned in \autoref{sec:GWsignal}, which depends on multiple parameters (\autoref{tab:parameters}), should reach an SNR of $2$ within 1 year of atom-interferometer measurement just before RLOF. 
The population of merging WDB computed above and displayed in \autoref{fig:WDBweighted}, which is assumed to be isotropic and randomly oriented, depends on the mass combination and distance.
An SNR distribution dependent on the same parameters, luminosity distance $d_L$ and masses $M_1$ and $M_2$, is required to compute the detectable WDB merger population.
This distribution can be found by averaging over the orientation and sky position parameters.
We already discussed the dependence of SNR on luminosity distance $d_L$.
Furthermore, we tested numerically with multiple realisations that the SNR distribution is the same for each $M_1$ and $M_2$, up to a normalisation given by the maximum SNR a given mass combination.

Armed with the WDB merger population synthesis and total merger rate from \autoref{sec:WDBpop}, the observed SN Ia rate from \autoref{sec:SNIarate}, the portion of binaries that are expected to go SN Ia from \autoref{sec:SNformation} and the SNR distribution depending on $d_L$, $M_1$ and $M_2$ from this segment, we can infer the detectable WDB merger rates with both approaches, and also the ones that are expected to go SN Ia.
The different WDB mass contributions to the detectable WDB mergers with SN Ia detonation are shown in \autoref{app:contributions}, while the resulting rates are presented below.

In addition, \autoref{app:corrections} presents several approximations made in this work and for the evaluation of the merger rates, and reports their validity and limitations.

\subsection{Results}\label{sec:results}

The results for the total merger rate and the  merger rate detectable with MAGIS Space, as well as the detection rate of binaries that detonate in SN Ia, are shown in \autoref{fig:Mergerrates}. 
The plots display the results in two formats: the cumulative rates (bottom plots), representing the total number of occurrences per year within a distance (as a function of that distance), and the differential rates (top plots), which show the contribution at each distance, corresponding to the derivative of the cumulative rates with respect to distance.
In addition, the right and left plots make the comparison between the outcome obtained with the two approaches: inferring the WDB merging population from the theoretical model of star formation history (left plots), and the observation of SN Ia (right plots). 
In the latter case, the results are scaled with the variable $DD\%$, which represents the percentage of SN Ia originating from the DD scenario and adds a layer of uncertainty.
In addition, the differential rate plots report both the results of the fiducial case and the uncertainties. For the cumulative figures, only the wider uncertainty bands are shown.

From the differential rate plots of \autoref{fig:Mergerrates} it can be seen how the total number of expected detections with MAGIS Space grows together with the distance up to $20-30$\,Mpc. 
At this distance, the differential rates of detectable events are maximal, signalling a peak in observations of both WDB mergers and coalescences that go SN Ia. 
Beyond this distance, there is a relatively quick drop for all observable quantities. 
However, for certain models, we might expect a relevant contribution to the detectable WDB merger by MAGIS Space up to a few hundred Mpc. 
This outcome arises from the most massive WDBs, which are expected to be louder and detectable up to higher distances, and not generate SN.

\begin{table}[t]
\noindent\makebox[\linewidth][c]{
    \centering
    \begin{tabular}{|c|c|c|c|}
        \hline
        \multicolumn{2}{|c|}{\textbf{Total volumetric rates}}   & \textbf{Model} &  \textbf{SN Ia rate} ($\times DD\%$)\\
        \hline\hline
         \multirow{2}{*}{\makecell{SN Ia rates from DD\\$[\text{yr}^{-1}\text{Mpc}^{-3}]$} } 
         & fiducial  & $(0.44-4.07)\times10^{-5}$ & $2.43\times10^{-5}$ \\\cline{2-4}
         & uncertainty & $(0.21-9.50)\times10^{-5}$  & $(2.11-3.53)\times10^{-5}$\\\hline
         \multirow{2}{*}{\makecell{WDB merger rates\\$[\text{yr}^{-1}\text{Mpc}^{-3}]$}} 
         & fiducial  & $8.11\times 10^{-5}$ & $(4.84-44.84)\times10^{-5}$\\\cline{2-4}
         & uncertainty & $(2.36-15.46)\times 10^{-5}$   & $(3.39-66.64)\times10^{-5}$\\\hline       
        \multicolumn{4}{c}{ }\\     
        \hline
        \multicolumn{2}{|c|}{\textbf{MAGIS Space rates}}  & \textbf{Model} &  \textbf{SN Ia rate} ($\times DD\%$)\\
        \hline\hline
         \multirow{2}{*}{\makecell{Detectable WDB\\merger rates $[\text{yr}^{-1}]$}} 
         & fiducial & $0.56$ & $0.34-3.12$ \\\cline{2-4}
         & uncertainty & $0.56-3.41$   & $0.29-24.41$\\\hline
         \multirow{2}{*}{\makecell{Detectable WDB merger rates\\that go SN Ia $[\text{yr}^{-1}]$}}
         & fiducial  & $0.25-0.56$ & $0.33-1.36$\\\cline{2-4}
         & uncertainty & $0.15-1.88$  & $0.29-2.67$ \\\hline    
        \multicolumn{4}{c}{ }\\
        \hline
        \multicolumn{2}{|c|}{\textbf{AEDGE rates}} &  \textbf{Model} & \textbf{SN Ia rate} ($\times DD\%$)  \\
        \hline\hline
         \multirow{2}{*}{\makecell{Detectable WDB\\merger rates $[\text{yr}^{-1}]$}} 
         & fiducial & $7.71\times 10^{2}$ & $(4.60-42.63)\times 10^{2}$ \\\cline{2-4}
         & uncertainty & $(7.71-92.30)\times 10^{2}$  & $(3.95-691.18)\times 10^{2}$ \\\hline
         \multirow{2}{*}{\makecell{Detectable WDB merger rates\\that go SN Ia $[\text{yr}^{-1}]$}}
         & fiducial  & $(2.78-7.49)\times 10^{2}$ & $(4.48-15.39)\times 10^{2}$\\\cline{2-4}
         & uncertainty & $(2.66-42.46)\times 10^{2}$   & $(3.86-66.73)\times 10^{2}$\\\hline
    \end{tabular}

}
    \caption{Results in terms of rates over the whole sky. The three tables show the values for the general volumetric rates over the all sky, the expected detections with MAGIS Space and also AEDGE. All outcomes are shown for both methods, from the theoretical model and starting from the SN Ia rates. Clearly, the latter method has an additional layer of uncertainty due to the percentage of SN Ia generated from the double degenerate scenario $DD$. In addition, both fiducial results and uncertainty due to the furthest results obtained from the different combinations of models and methods are shown.}
    \label{tab:finaltable}
\end{table}

The total yearly rates over the whole sky are reported in detail in \autoref{tab:finaltable}. 
The total SN Ia rates and total WDB merger rates reported in \autoref{fig:Mergerrates} and shown in the section ``Total volumetric rates'' of the table are independent of the detector. 
The outcomes of this analysis exhibit a good agreement between the two computational methods: the theoretical model approach on population synthesis and the approach based on observed SN Ia rates.
Both results indicate that MAGIS Space is expected to detect at least a WDB merger every $3$ years, one of which may result in a SN Ia approximately every $4$ years, and quite possibly more events than this. 
The SN Ia approach spans its results over a larger range of rates and suggests higher values. 
Nevertheless, the increased rates of this approach agree with the expectation that the DD scenario does not explain the entirety of SN Ia events, but only a portion.

As previously mentioned, WDB mergers are considered detectable if they reach an SNR of $2$ within $1$ year of observation. 
Having a detector running for a longer time would increase the expected detection rate. 
We are only showing the rate of detecting actual mergers, many more WDBs close to merger could be measured, both at larger distances and at lower masses. 
In addition, we can recall that our SNR computation is stopped at RLOF. 
This cutoff represents a conservative assumption regarding the loudness of the signal, and hence, these are conservative predictions for the detection rates. 
As described in \autoref{sec:MT} the GW signal is expected to continue and become louder after RLOF.

The same calculations done for assessing the capabilities of MAGIS Space can also be applied to other detectors.
For example, AEDGE has a similar technology to MAGIS Space and is expected to cover approximately the same range of GW frequencies, but with an improved sensitivity.
We have performed an analogous analysis for AEDGE, considering a different sensitivity (see \autoref{fig:CharacteristicStrain}) but the same detector configuration. 
A comparison between MAGIS Space's and AEDGE's detector-dependent results for the model approach is displayed in \autoref{fig:FinalFigure} and described in detail in \autoref{tab:finaltable}.
The superior sensitivity of AEDGE leads to rates of detection roughly three orders of magnitude larger than for MAGIS Space. 
In fact, the fiducial results for MAGIS Space forecast a maximum of $3$ WDB merger detections per year.
With AEDGE, a minimum of a few hundred observations of mergers per year, which could also go SN Ia, are expected. 
This difference is highlighted in \autoref{fig:FinalFigure}, which shows how the latter detector can observe binaries as far as a few thousand Mpc.

\subsubsection{SN Ia non-detection}\label{sec:noSNIa}

We would also like to consider whether we can confirm the SD scenario applies to certain SN Ia.  If a SN Ia is detected electromagnetically and we do not see a corresponding gravitational wave source, we could potentially confirm the SD scenario.  However this only works if we know that we should have seen the GW's had the SN been caused by a WDB merger.

\begin{figure}[t]
\centering
\includegraphics[height=0.7\linewidth]{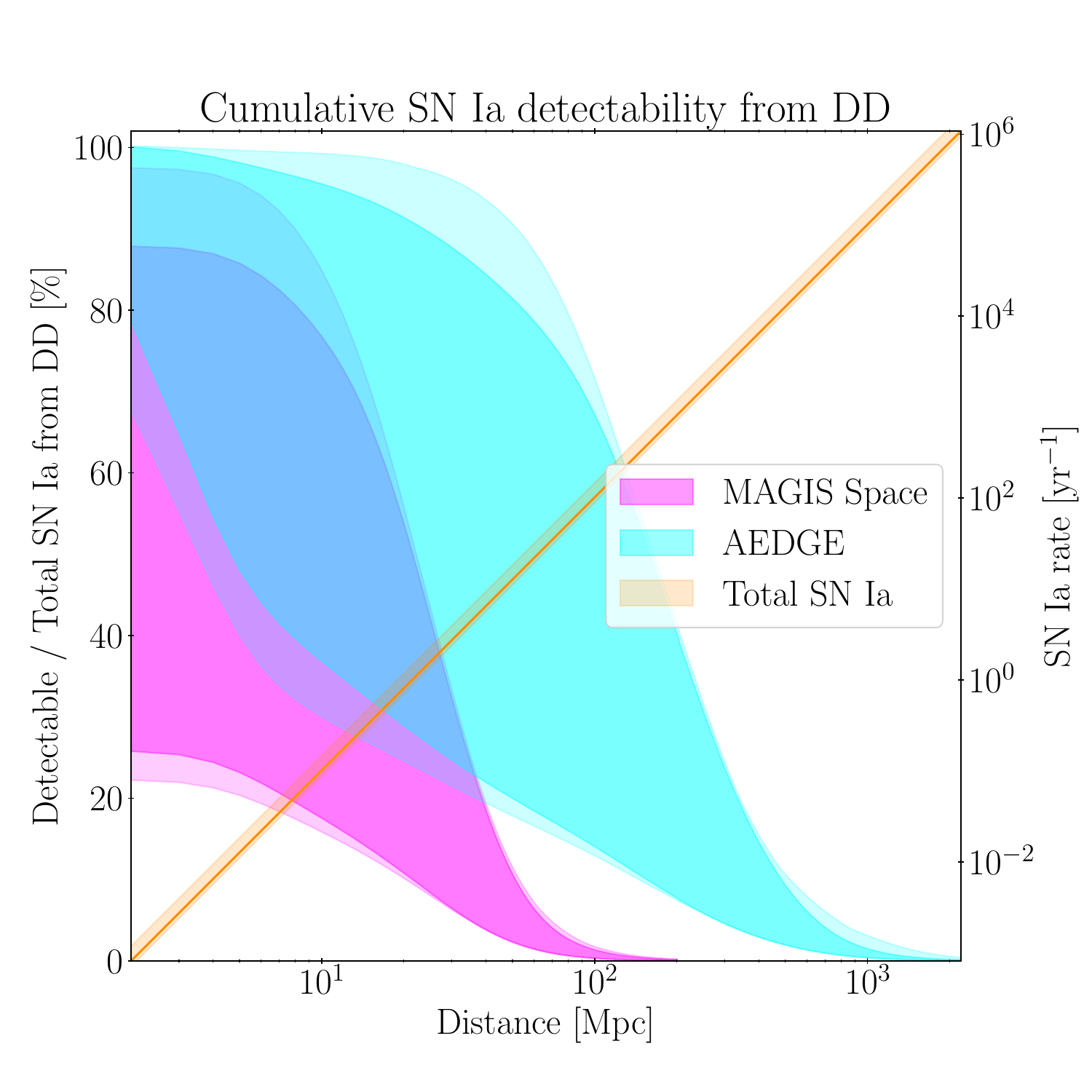}
\caption{The cumulative GW detectability of SN Ia coming from the DD scenario, hence from WDB merger, is shown for both the MAGIS (pink) Space and AEDGE (cyan) instruments, and refers to the left axis of the plot. We plot the findings using the theoretical model approach. 
The bands are the result of considering both extreme SN Ia production scenarios (see \autoref{sec:SNformation}). 
As for \autoref{fig:Mergerrates}, the darker colour represent the results coming from the fiducial star formation model $\gamma=1.75$, $\alpha=4$ from \autoref{sec:WD}, while the fainter ones represent the uncertainty due to other models. 
Additionally, the total number of SN Ia events (from any scenario), computed with equation~\eqref{eq:SNIarate}, within a certain luminosity distance is plotted, as indicated on the right axis. Again, the solid line indicates the fiducial value and the colour band the uncertainty.  
}
\label{fig:noSNIa}
\end{figure}

From the previous result, we can directly compute the fraction of SN Ia originating from the DD scenario that can be expected to be detected through GW. \autoref{fig:noSNIa} illustrates these findings, for the theoretical model approach, in comparison to the total number of SN Ia that might be seen with EM telescopes.
The proportion of detectable GW signals from WDBs that result in SN Ia, corresponding to the DD scenario, covers a broad range of values.
The upper bound, for both detectors, arises from the conservative SN Ia formation case discussed in \autoref{sec:SNformation}. In this regime, WDBs are much more massive (see \autoref{fig:WDBweighted}), hence loud, leading to a larger portion of detectable GW signal, which leads to SN Ia. More specifically, MAGIS Space is expected to detect more than $75\,\%$ of SN Ia from the DD formation channel within $10$\,Mpc, while AEDGE more than $95\,\%$. 
The lower bound is determined by the most optimistic assumptions about the SN Ia formation scenario, as it considers much lower masses and fainter WDBs, which are additionally more abundant\footnote{Most WDBs that go SN Ia in the conservative case have individual WD masses around $0.6\,M_\odot$, see \autoref{fig:WDBweighted}.}. MAGIS Space is expected to detect a minimum of $15\,\%$ of the GW signature of SN Ia from the DD channel within the same distance as above, $10$\,Mpc. As we look to larger distances, these fractions decline sharply. Inside such a limited volume, only a limited number of visible SN Ia are expected, approximately one every $10$\,years. Therefore, a lack of GW signal from visible SN Ia events in this region would favour the SD formation channel only after a century.
Conversely, AEDGE provides significantly stronger constraining power. While the lowest percentage of GW detectable SN Ia within $10$\,Mpc only doubles for AEDGE compared to MAGIS Space, cutting in half the observation period needed for a significant result, a much larger volume can be considered with AEDGE.
For $100$\,Mpc, the forecasted visible SN Ia are on the order of a hundred per year, and the fraction of measurable GW signatures from WDB that may go SN Ia with AEDGE spans between $13\,\%$ and $71\,\%$. Consequently,  SN Ia observation in the EM spectrum without a GW counterpart measured by AEDGE will quickly set strong constraints on the DD formation channel of SN Ia, favouring SD.

Null-detections of GW emissions from SN Ia in a year by AEDGE will constrain the contribution of the DD formation channel of SN Ia to less than $10\,\%$.
On the other hand, if SN Ia arise from a roughly even mixture of SD and DD, we could tell that DD was contributing a part of the total, but it would be difficult to conclude that SD was a contributing part, given the large uncertainties on the detectable fraction in \autoref{fig:noSNIa}, even for AEDGE.

\section{Discussion} \label{sec:Discussion}

In this work, we explore the capabilities of MAGIS Space and AEDGE, see \autoref{sec:Detector} and \autoref{fig:CharacteristicStrain}, for detecting white dwarf binary (WDB) mergers. 
In \autoref{sec:WDBevo}, we discussed many properties of WDBs in order to better understand their evolution and final stages. 
We then analysed their gravitational wave (GW) signatures using the tools described in \autoref{sec:GW}.
A strategy to recognise an imminent merger with GW measurement and observe the subsequent electromagnetic (EM) counterpart for each WDB is described in \autoref{sec:Mergersignal}.
In \autoref{sec:mergerrate} we made predictions for the detectable merger rates and Supernova Type Ia (SN Ia) observations with atom-interferometers.

\begin{figure}[t]
    \centering
    \includegraphics[width=0.8\linewidth]{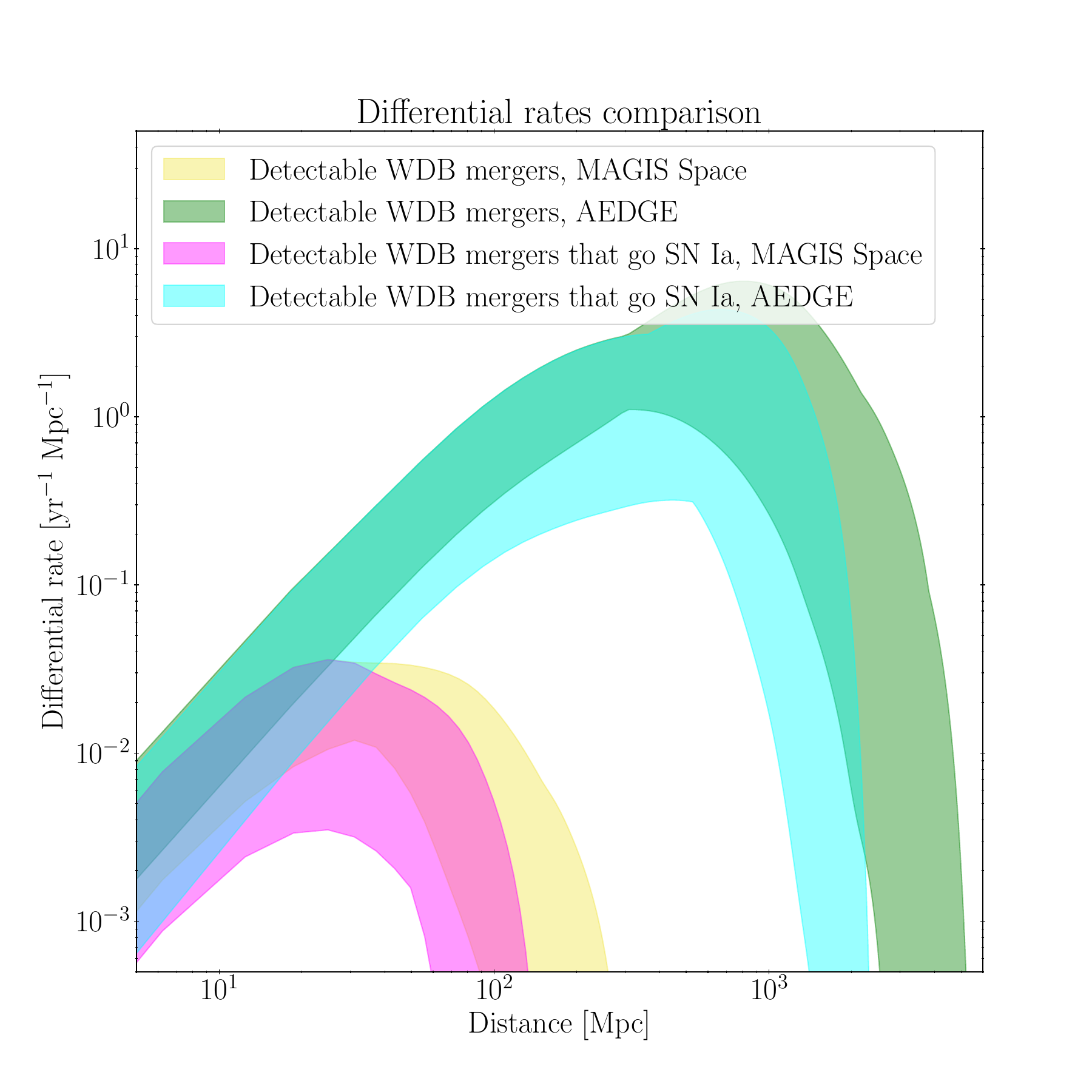}
    \caption{Differential rates on the detectable WDB mergers and their possibility of exploding in a SN Ia for MAGIS Space detector and AEDGE. The plot shows the uncertainty bands resulting from the merger rate analysis that uses the total WDB merger population from theoretical models described in \autoref{sec:mergerrate}. A detailed breakdown and values of the total rates are displayed in \autoref{tab:finaltable}.}
    \label{fig:FinalFigure}
\end{figure}

We characterized the GW signal before merger and determined that MAGIS Space will be able to estimate accurately the angular localisation $\Omega$, with precision between $\sim1-10^{-3}\,{\rm deg}^2$ at $25$\,Mpc, and chirp mass $\mathcal{M}_c$, better than $10^{-5}\,M_\odot$, of the observed binaries (Figures \ref{fig:MeasurementPrecision_dL_Sky} and \ref{fig:MeasurementPrecision_q_Mc}).  The measurements of the mass ratio between the two companions $q$ and their luminosity distance $d_L$ will have much larger uncertainties.
In addition, we studied the possibility of observing WDB mergers and their aftermath with EM telescopes.
The high precision in angular localisation enables us to accurately determine the binary’s position in the sky.

We have developed a strategy to either predict or promptly identify the merger.
Most binaries, with $M_1\lesssim 1.2\,M_\odot$, undergo a period of slow mass transfer
before merging, which alters the WDB evolution and generates a change to the phase of the GW waveform (\autoref{fig:RLOF1pi}).
Depending on the binary's distance, thus the loudness of its GW signal, this dephasing can be observed and represents an early warning for the coalescence, leading to an opportunity for multi-messenger astronomy.
By contrast, very massive binaries are driven by GW radiation until merger and do not produce an early warning due to mass transfer. 
Although a merger cannot be predicted in that case, it can be quickly identified due to the disappearance of the loud GW signal of massive WDBs. 
Differentiating between the two cases with observations and anticipating the right approach for merger detection is possible due to the high precision in measuring $\mathcal{M}_c$.
A final SN Ia event could potentially be detected by either technique, as its post-detonation luminosity persists for several weeks.

Regarding the observable WDB merger rates and their possibility to go SN Ia, we performed the calculations using two different approaches, which converged on the same outcome.
The first method is based on a theoretical WDB merger population model and provides us with results without free variables. 
The second method is based on the same WDB mass function, but refers to observed SN Ia rates by adding the double degenerate (DD) scenario contribution to SN Ia as a variable. 
According to our estimates, MAGIS Space is expected to detect at least one WDB merger every $3$ years and up to $3$ coalescences annually, under the fiducial model considered.
Most of these WDB mergers are expected to lead to a SN Ia event, decreasing the probability of this occurrence only to a range of $0.25-1.36$ detections per year.

We performed analogous computations for the sibling project AEDGE, which is projected to have an improved GW sensitivity. 
We estimated that AEDGE would observe more than a hundred WDB mergers each year with an EM counterpart of interest. 
The differential detectable WDB merger rates over distance, given by the theoretical model approach, are shown in \autoref{fig:FinalFigure} (see also \autoref{tab:finaltable} and \autoref{fig:Mergerrates}). 
These results are based on $1$ year of observation prior to the start of mass transfer.
Longer measurements would allow both instruments to detect more merging binaries.
As discussed throughout the text, we have made a number of simplifying assumptions in our calculations, see also the discussion in \autoref{app:corrections}.
\\[1\baselineskip]
Observations of white dwarf binaries by MAGIS Space or AEDGE are expected to have a wide range of applications, revolutionising multi-messenger astrophysics.
Current models of the end stage of WDBs are highly uncertain.  Observations of these events in both the gravitational and electromagnetic spectra will provide valuable information about these models.
SN Ia are one of the most interesting end-stage scenarios.
The observation of this event would represent a confirmation that the DD formation channel is responsible for at least some of the SN Ia (see \autoref{sec:WDBevo}), shedding light on this detonation.
Moreover, such a multi-messenger observation of WDB mergers could offer valuable clues on the Hubble puzzle by helping inform our understanding of SN Ia, which are used as standard candles, and providing complementary measurements of distance from the same source.
The absence of a corresponding SN Ia detection in either the EM or GW channels would, however, be equally important for determining the origins of SN Ia and the characteristics of WDBs.
On the one hand, while MAGIS Space provides limited constraining power, if AEDGE fails to detect any GW counterpart for the observed SN Ia in the EM spectrum, the SD scenario would be strongly favoured and must be responsible for at least a large part of the SN Ia (see \autoref{fig:noSNIa}).
On the other hand, alternative final events could take place instead of SN Ia and be observed: a collapse into a neutron star or black hole or a different type of detonation.
The recording of the GW signature of the mass transfer period will then play a fundamental role, not only in giving us novel and unique insights into the underlying dynamics and composition of white dwarfs, but also in determining the conditions for the various WDB merger outcomes.

By observing white dwarf binaries in the mid-band with gravitational wave detectors, we will learn more about the origin and properties of SN Ia and about the complicated dynamics of these binary mergers, even when they do not trigger a supernova.

\acknowledgments
We thank Ariel Goobar, Roger Romani, Seppe Staelens, and Bob Wagoner for insightful discussions. 
GS and SB acknowledge support from the DFG under grant 396021762 - TRR 257: Particle Physics Phenomenology after the Higgs Discovery. 
This work was supported in part by NSF Grant No.~PHY-2310429, Simons Investigator Award No.~824870,  the Gordon and Betty Moore Foundation Grant No.~GBMF7946, and the John Templeton Foundation Award No.~63595.


\appendix
\section{Merger rates contribution} \label{app:contributions}
\begin{figure}[t]
\centering
\includegraphics[width=\linewidth]{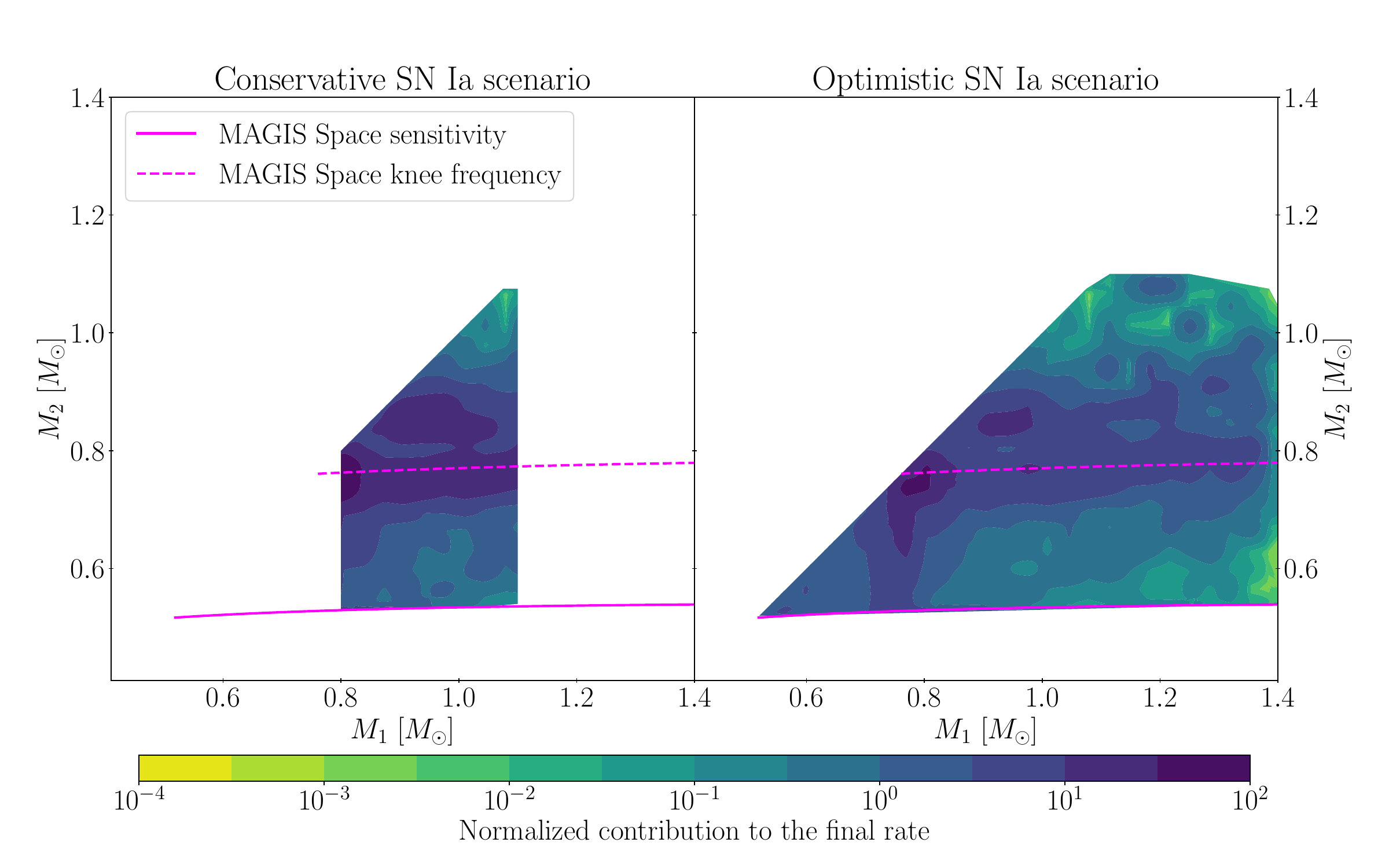}
    \caption{The two plots show the normalised density contributions of different binary mass combinations to the total detectable merger rate of WDB mergers that explode in SN Ia (in units of $M_\odot^{-2}$). 
    The conservative \eqref{eq:SNconservative} (left) and optimistic \eqref{eq:SNoptimistic} (right) SN Ia production cases are reported. 
    For reference, the lowest frequency sensitivity of MAGIS Space is displayed, as well as the threshold at which the sensitivity starts to decrease towards lower frequency (see \autoref{fig:CharacteristicStrain}).
    We apply Gaussian smoothing to erase numerical noise.}
    \label{fig:Mergerratescontributions}
\end{figure}

\autoref{fig:Mergerratescontributions} shows the contributions to the total detectable signal of WDB mergers measured by MAGIS Space, that are expected to go SN Ia. 
We studied the fiducial model considering the SN Ia rate approach, for which all SN Ia are due to the DD scenario.
As discussed in \autoref{sec:WDBdetections}, for the WDB population synthesis, we used the star formation history, based on the different metallicities, from~\cite{Chruslinska2019}, and the evolution model $\gamma\alpha$, where $\gamma=1.75$ and $\alpha=4$, following~\cite{Seppe24,Hofman24}. 
The total detectable WDB mergers that end in SN Ia are reported in \autoref{tab:finaltable}: $0.33-1.36$ per year.
Both conservative and optimistic SN Ia production cases from \autoref{sec:SNformation} are analysed and shown in the plots.

Starting from the top of \autoref{fig:Mergerratescontributions}, it can be seen that the contributions grow with decreasing masses, because of the larger population of lighter WDB mergers, as shown in \autoref{fig:WDBweighted}. 
However, the contribution starts to decrease once we pass the threshold given by MAGIS Space's ``knee frequency'', after which, in addition to the binaries being less loud, the interferometer sensitivity decreases (see \autoref{fig:CharacteristicStrain}).
In both cases, the largest contributions to the detectable WDB mergers that go SN Ia come from binaries with similar masses, around $0.8\,M_\odot$ each.

\section{Corrections}\label{app:corrections}
Many assumptions made in this project's analysis were conservative. Nevertheless, a few effects that might influence the results negatively were not considered and must be mentioned. 

Firstly, gravitational wave signals will redshift. As a consequence, for large distances, the frequencies at the detector will be significantly lower than at the source. This difference could influence the SNR and detectability of certain binaries and modify our outcomes. We inspected the magnitude of this effect for the two considered detectors, MAGIS Space and AEDGE. 
Our investigation showed that the major loss of detections does not come from the lowest frequency sensitivity of the instruments, but from the frequency at which the sensitivity starts dropping quickly, instead, also called the knee frequency (see \autoref{fig:CharacteristicStrain}). 
Moreover, the corrections to the detection rates expected for MAGIS Space are of the order of $\sim1\,\%$, while AEDGE could lose up to $\sim50\,\%$ of its observations close to the merger. 
As expected, this effect is more relevant for AEDGE as it can measure WDB to much larger distances, whose signal experiences a larger redshift. The small correction on MAGIS Space detections confirms our results and warns us about this effect for the AEDGE instrument. 
Nonetheless, the expected detectable WDB mergers with AEDGE remain higher than $100$ per year. 

Another issue regards the WDB merging population synthesis. 
As previously mentioned in \autoref{sec:WDBpop}, we considered binaries reaching RLOF as merging, thus neglecting any effect between that moment and the actual merger. 
Binary stability might indeed vary the merging population today, especially for the detections with AEDGE.
The consequences on our final rates of these effects were not studied, due to the high uncertainty on binary stability.
However, extreme stability only affects low mass WDB, which are not as interesting for our conclusions, both in terms of loudness and final event (SN Ia). 
It is worth noting that in some cases (such as this one), the approximation used in this work leads, counterintuitively, to a conservative result. 
For example, considering only stable binaries and starting from the observable SN Ia rate would mean having more massive, hence detectable, WDBs which generate SN Ia.

A comment on the fraction of WDB mergers that generate SN Ia seems also needed. 
In \autoref{sec:SNformation} we selected two WDB mass ranges that might produce these events (see \autoref{fig:WDBweighted}) and estimate that the real scenario will fall somewhere between these two cases. 
Selecting these mass regions has mainly two effects: the first is that, together with the weight of the DD scenario in producing SN Ia, this assumption determines the total number of WDB mergers for the SN Ia rate approach; the second is to determine the detectable WDB merger rate that could go SN Ia.
The first effect can be used to check our outcome.
It can be seen in \autoref{fig:Mergerrates} and \autoref{tab:finaltable} that the selected SN Ia progenitors lead to an agreement between the results from the model and SN Ia rate methods for a DD scenario contribution of $\sim25-50\,\%$.  
However, the same matching can be achieved with different combinations of DD scenario contribution to SN Ia events and WDB mass regions that detonate in SN Ia\footnote{E.g. a stricter condition on the WDB masses for generating SN Ia would match the total number of mergers from the model method if the DD scenario explains the totality of these events.}.

Regarding the dynamics close to a merger, in our project, we studied a simplified version of binary evolution. Many effects, such as tidal interactions~\cite{Piro:2011qe, Lai:2011xp, Fuller:2014ika, McNeill:2019rct} were neglected. 
A more careful analysis, including side effects, in addition to being beyond the goal of our work, is not expected to improve the results. 
In fact, there are large uncertainties on the late stage dynamics of WDB, e.g. the spin of WDs in binaries, the actual start of mass transfer, the stability of the binaries, \dots
We assume that the effect of secondary corrections, such as tidal interactions, would be irrelevant compared to the uncertainties.

Next, we considered the change in star formation rate density and metallicities at different distances for the population synthesis, but we then required the condition that all WDB must be merging today. 
Different redshifts of the GW sources could have a slightly different WDB merging population due to shorter time available to them to reach RLOF.
This difference is expected to grow as we look further away. 
As a consequence, once again, this effect is more relevant in the case of AEDGE, while it is reasonable to neglect it when considering MAGIS Space.

Finally, we did not consider any signal detection challenge. 
As mentioned, for example, in~\cite{Seppe24,Hofman24}, many WDBs, but also other sources, will be measured by atom-interferometers. Disentangling the various signals and recognising the WDB merging one is expected to be an engaging task, especially in the case of AEDGE sensitivity, which will gather many more inputs. 
We only mentioned in \autoref{sec:parameters} that we do not consider the Doppler shift caused by Earth's rotation, as we expect this effect to be easily removed from the signal.

The focus of this project is characterising MAGIS Space's future observations on WDB. 
All approximations applied in this project are expected to be robust for this detector.
The results computed with AEDGE rely on the same assumptions used for its sibling detector but may not fully hold in the AEDGE case, and should therefore be interpreted with some caution.

\bibliographystyle{JHEP.bst}
\bibliography{TheBib}

\end{document}